\documentclass[a4paper,11pt]{article}

\usepackage[utf8]{inputenc}
\usepackage[british]{babel}
\usepackage{csquotes}

\usepackage[
backend=biber,
style=numeric-comp, 
sorting=none 
]{biblatex}
\usepackage{debug}

\newcommand{\COMMENTOOK}[1]{}
\usepackage{graphicx}
\usepackage{subcaption} 
\usepackage{color} 

\usepackage{authblk}
\usepackage{hyperref}
\usepackage{amsmath}
\usepackage{amssymb}
\usepackage{amsfonts}
\usepackage{enumitem}
\numberwithin{equation}{section}

\newcommand{\eom}{{e.o.m}}
\newcommand{\wrt}{{w.r.t.}}
\newcommand{\bcs}{{boundary conditions}}

\newcommand{\oh}{ \frac{1}{2} }

\newcommand{\C}{ \mathbb{C} }
\newcommand{\N}{ \mathbb{N} }
\newcommand{\R}{ \mathbb{R} }

\newcommand{\cA}{{\cal A} }
\newcommand{\cH}{{\cal H} }
\newcommand{\cM}{{\cal M} }
\newcommand{\cN}{{\cal N} }
\newcommand{\cX}{{\cal X} }
\newcommand{\cZ}{{\cal Z} }

\newcommand{\BO}{Boost Orbifold}
\newcommand{\NSO}{Null Shift Orbifold}

\newcommand{\tz}{ t_0 }

\newcommand{\ksi}[2]{ | #1 \{#2\} \rangle }
\newcommand{\ksiS}[4]{ | #1 \{#2\}; #3, #4 \rangle_S }
\newcommand{\kcohsi}[2]{ | #1 \{#2\} ) }
\newcommand{\kcohsiS}[4]{ | #1 \{#2\}; #3, #4 )_S }
\newcommand{\bsi}[2]{ \langle #1 \{#2\} | }

\newcommand{\bcohsi}[2]{ ( #1 \{#2\} | }
\newcommand{\bcohsiS}[4]{ {}_S( #1 \{#2\}; #3, #4| }

\newcommand{\HH}{Hamiltonian in  Heisenberg picture }
\newcommand{\Scp}{{Schroedinger picture} }

\newcommand{\pii}{ p_{(i)} }
\newcommand{\Ri}{ R_{(i)} }

\newcommand{\bs}[1]{ {s^{[#1]}}{} }
\newcommand{\bc}[1]{ {c^{[#1]}}{} }

\title{
On the breakdown of the perturbative interaction picture in Big Crunch/Big Bang
or\\
the true reason why perturbative string amplitudes on temporal orbifolds diverge
}

\author{Igor Pesando}
\affil
{%
  Dipartimento di Fisica, Universit\`{a} di Torino \authorcr
  and I.N.F.N. -- sezione di Torino \authorcr
  Via P.\ Giuria 1, I-10125 Torino, Italy
}

\begin{document}
 
\maketitle

\begin{abstract}
    We discuss how the perturbative particle paradigm fails in certain
    background with space-like singularity
    but asymptotically flat which should admit a S-matrix.
    
    The Feynman approach relies on the interaction picture.
    This approach means that we can interpret interactions as
    exchanges of particles. Particles are the modes of the
    quadratic part of the Lagrangian.  In certain backgrounds with
    space-like singularity the interaction Hamiltonian is well defined
    but the perturbative expansion of the evolution operator through
    the singularity and the perturbative $S$ matrix do not exist.

    On the other hand, relying on minisuperspace approximation 
    we argue that the non perturbative evolution operator does exist.

    The complete breakdown of the perturbative expansion
    explains why the perturbative computations in the covariant
    formalism in string theory in temporal orbifold fail, at least at
    the tree level.

\end{abstract}

\section{Introduction}

While this paper is mostly on QFT and its behavior on singular
spacetimes describing some models of Big Crunch/Big Bang
its reason has roots in string theory.
String theory, as a promising candidate for a theory of quantum
gravity, is supposed to provide a satisfactory description of Big
Bang/Big Crunch type singularities, or at least a S matrix in
asymptotically flat spaces.

We want therefore to construct and study
stringy toy models capable of reproducing a
space-like (or null) singularity which appears in 
space at a specific value of the time coordinate and then disappears.

The easiest way to do so is by generating singularities by
quotienting Minkowski with a discrete group with fixed points,
i.e. orbifolding Minkowski.
In this way it is possible to produce both space-like singularities
and supersymmetric null singularities
\cite{Horowitz:1989bv,Horowitz:1990sr,Nekrasov:2002kf,Craps:2002ii,
Liu:2002ft,Liu:2002kb,
Fabinger:2002kr,David:2003vn,Craps:2008bv,Madhu:2009jh,Narayan:2009pu,
Narayan:2010rm,Craps:2011sp,Craps:2013qoa} (see also
\cite{Cornalba:2003kd,Craps:2006yb,Berkooz:2007nm} for some reviews).
Another possible way which is a generalization of the previous
orbifolds with null singularity is consider gravitational shock wave
backgrounds \cite{deVega:1990kk,deVega:1990ke,deVega:1990gq,deVega:1991nm,Jofre:1993hd,Kiritsis:1993jk,DAppollonio:2003zow}.

It happens that in these orbifolds
the four tachyon closed string amplitude diverges in
some kinematical ranges, more explicitly for the \NSO{} (which may be
made supersymmetric and has a null singularity) we have
\begin{equation}
\cA^{(closed)}_{4 T}
\sim
\int_{q\sim \infty} \frac{d q}{|q|} q^{ 4- \alpha' \vec p^2_{\perp\, t}}
,
\end{equation}
so the amplitude diverges for $\alpha' \vec p^2_{\perp\, t}<4$ where
$\vec p_{\perp\, t}$ is the orbifold transverse momentum in $t$ channel.
Until recently this pathological behavior has been interpreted in the
literature as “the result of a large gravitational backreaction of the
incoming matter into the singularity due to the exchange of a single
graviton”.
This is not very promising for a theory which should tame quantum
gravity.

What has gone unnoticed is that if we perform an analogous computation
for the four point open string function we find
\begin{equation}
\cA^{(open)}_{4 T}
\sim
\int_{q\sim \infty} \frac{d q}{|q|} q^{ 1- \alpha' \vec p^2_{\perp\, t}}
tr\left(\{T_1, T_2\} \{T_3, T_4\}\right)
,
\end{equation}
which is also divergent when for $\alpha' \vec p^2_{\perp\, t}<1$
(\cite{Arduino:2020axy,Hikida:2005ec}).
This casts doubts on the backreaction as main explanation since we are
dealing with open string at tree level.
This is further strengthened by the fact that three point amplitudes
with massive states may diverge \cite{Arduino:2020axy}
when appropriate polarizations are chosen.
For example for the three point function of two tachyons and
the first level massive state we find for an
appropriate massive string polarization
\begin{equation}
\cA^{(open)}_{T T M}
\sim
\int_{u\sim 0} \frac{d u}{|u|^{5/2}}  tr\left(\{T_1, T_2\} T_3\right)
.
\end{equation}
In \cite{Arduino:2020axy} this was interpreted as a non existence of the
underlying effective theory.
We now revisit this assertion and argue that the effective theory does
exist but the usual approach based on the perturbative expansion in
the interaction picture completely breaks down.

In this paper we consider what happens when we use perturbation theory
in a time dependent background with a space singularity.
It is somewhat obvious that we do  not expect to find a well behaved
perturbation theory because of the singularity.
One could expect some kind of pathology like the series being
asymptotics.
We find a much worse behavior: a complete breakdown of perturbation
theory, i.e. perturbation theory does  not exist.
Let us be more precise.
We consider as unperturbed theory the free, non interacting QFT in the
given singular time dependent background and then add interactions.
We then use the usual interaction picture approach.
This approach when used perturbatively naturally leads to
Feynman diagrams and a nice particle interpretation of interactions.
In the backgrounds we consider all of this suffers from a complete
breakdown.
There is no perturbative expansion in the usual sense.
This prompts the question whether it is perturbation theory which
fails or it is the very interacting theory which does not exists.
To answer this question we consider the minisuperspace approach, i.e.
the consider the QFT reduced to the spacially homogeneous
configurations (see \cite{Halliwell:1989myn} for review).
In this limit the theory reduces to Quantum Mechanics.
We then show that these models do exist.
One could wonder whether this reduction is a big limitations and the
answer is no since it has been
shown \cite{Arduino:2020axy,Craps:2013qoa} that the troubles in
perturbation theory stem from these configurations.
The main difference with the work from the 80s and 90s is that we are
interested in going through the singularity and not giving the
boundary conditions at the Big Bang.

This result stresses the importance of treating some sectors as exactly
as possible in order to get a perturbation theory for the remaining sectors.
Even so we are left with the unanswered question whether it is really
consistent to treat QFT on a given singular background without
considering the backreaction. It is somewhat likely that
the gravitational background and
the matter should evolve together, especially in a background
which has space singularities.
Given the results of this paper it could be sufficient to consider the
minisuperspace approximation to get a reasonable approximation.
In any case this route is fraught with subtleties like the ``problem
of time'' (see \cite{Isham:1992ms} for a review).

The paper is organized as follows.

In section \ref{sect:backgroud} we discuss the background of interest,
the generalized Kasner metrics (of which the \BO{} is a very special case)
and the simplest interacting field theory, i.e. the scalar field and
its minisuperspace approximation.

In section \ref{sect:simplest_example} we discuss the simplest example
where the perturbative interaction picture breaks completely down: the
time dependent harmonic oscillator with
$\Omega^2(t) = \omega^2 + \frac{k}{t^2}$ and $k\le \frac{1}{4}$ so
that $\Omega^2$ may become negative.
While this model is natural since it corresponds to, for example, de
Sitter modes in conformal time the splitting we perform between the
unperturbed Hamiltonian and the perturbative part is somewhat artificial
but it is chosen in order to get the simplest example as possible.

In section \ref{sect:interacting_qm_model} we consider
the interacting theory and we show that generically the perturbation
theory of the interacting minisuperspace model does not exist.
We then study the minisuperspace model non perturbatively
and show that it does exist.
The model exhibits two different behaviors: either it is dominated by
the combination  of kinetic and interaction terms or it is dominated
by the interaction term alone.

Finally in section \ref{sect:string_on_temporal_orbifolds}
we discuss what this means for the divergences in string theory.
In nuce string theory is well, at least at tree level but the non
Hamiltonian perturbation theory has troubles.
Moreover we point out that the usual approach to orbifolds used in
string theory is not on very sound basis when temporal orbifolds are
considered since the orbifold generators are dynamical generators,
except for \NSO{} in light-cone gauge.

\section{The background}
\label{sect:backgroud}
Our starting point is to consider a class of backgrounds which have a
space-like singularity and on these
backgrounds write down the simplest interacting scalar theory.

Previous results from the analysis of issues in open
string amplitudes in these
backgrounds \cite{Arduino:2020axy,Craps:2013qoa}
hint toward the fact the all
troubles derive from special field configurations
to which we restrict.
In particular this means that we restrict these theories to space
independent but time dependent
fields in the space-like singularity case.

More precisely this paper we are going to consider the following
family of backgrounds.

\subsection{Kasner-like metrics}
The metric we consider is a generalization of the original Kasner
metric and reads
\begin{equation}
  d s^2
  = -d t^2 + \sum_{i=1}^{D-1} |t|^{2 \pii} \Ri^2 (d x^i)^2
  ,~~~~
  0\le x^i < 2\pi
  ,
  \label{eq:Kasner_metric}
\end{equation}
where we consider $t\in\R$ and not only $t>0$ and therefore we have
written $|t|$ since $\pii\in\R$.
We have also considered the $x^i$ to be compact in order to get a well
defined minisuperspace approximation of the scalar field as in
eq. \eqref{eq:qm_model_scalar_in_Kasner}.  

The original Kasner metric corresponds to the case where
$\sum_i \pii = \sum \pii^2 =1$ and space is not compact.
It requires that at least one $\pii$ is negative when at least two
$\pii$ are different from zero and corresponds to an empty space-time.
Another special case is when only $p_{(1)}=1$ and corresponds to Milner space.

All these metrics have a singularity at $t=0$
which is the target of our investigation.
They have generically also a singularity for $|t| \rightarrow \infty$
when some $p$ is negative.
When all $p$ are positive the metric requires repulsive matter.

For generic $\pii$ this metric is not a consistent string background
since $Ric \ne 0$.

\subsection{Interacting scalar models}
It is the immediate to write down the action for an interacting real
scalar field as
\begin{align}
  S
  =&
  \int d t \prod_i d x^i\, \prod_i \Ri |t|^{\sum_i \pii}
  \Biggl[
  \oh \dot \phi ^2
  -
  \oh \sum_i \frac{1}{\Ri^2 |t|^{2 \pii}} (\partial_i \phi)^2
  \nonumber\\
  &-
  \oh m^2 \phi^2
  -
  \frac{1}{n} g_n \phi^n
  \Biggl]
  ,~~~~
  n\in{4,6,\dots}
.
\end{align}
According to the analysis of string theory on \BO~
\cite{Arduino:2020axy,Craps:2013qoa}
the  problems for
this theory derive from the field configurations where the field
depends on time only.
Restricting to this configuration we get the quantum mechanical model
\begin{align}
  S
  =&
  \prod_i(2\pi \Ri)
  \int d t  |t|^{2 A}
  \Biggl[
  \oh \dot \phi ^2
  -
  \oh m^2 \phi^2
  -
  \frac{1}{n} g_n \phi^n
    \Biggl]
,
\label{eq:qm_model_scalar_in_Kasner}
\end{align}
where we have defined $2 A = \sum_i \pii$ for compactness.
We consider only the case where $A>0$.

\section{The simplest example of failure of the perturbative expansion
  in interaction picture: the time dependent harmonic oscillator}
  \label{sect:simplest_example}
In this section we would like to discuss how the usual perturbative
expansion in  interaction picture may completely break down
when the interaction Hamiltonian has time singularities.
This may happen despite the complete model is well defined.

In particular the model we want to consider is
\begin{equation}
  L_R= |t|^{2 A} \left( \oh \dot y ^2 - \oh \omega^2  y^2 \right)
  \label{eq:L_Rosen_time_dep_harm_osc}
  ,
\end{equation}
which corresponds to the non interacting scalar on Kasner metrics.
Two special cases are $A=0$ and $A=\oh$ and both correspond to the
flat space but in Minkowski and Milne (Boost orbifold) coordinates. 
Upon a change of coordinates as
\begin{equation}
  x = |t|^A y
  ,
\end{equation}
we  get
\begin{equation}
  L_B= \oh \dot x ^2 - \oh \left( \omega^2 + \frac{k}{t^2} \right) x^2
  + \frac{d}{d t}\left( \oh \frac{A}{t} x^2 \right)
  ,~~~~
  k=A (1-A)\in (-\infty,\frac{1}{4})
  \label{eq:L_Brinkmann_time_dep_harm_osc}
  .
\end{equation}
The total derivative is uninfluential at the classical level while at the
quantum it implies a relative time dependent phase for the wave function in the
two coordinate systems see eq. \eqref{eq:psiB_psiR_non_interacting}. 

Notice that when $k$ is negative ($A>1$ or $A<0$) the potential is unbounded
from below but despite this the full model is well defined.
That this may happen is not a surprise since the hydrogen atom exists.
On the other side in the flat space $A=0,\oh$ the potential is always
bounded from below.
In particular the $A=\oh$ case is the Milner space which is a subset of
Minkowski space and even so the model has a singular potential.

This model emerges 
besides the obvious case of the non interacting scalar in Kasner-like
metrics mentioned above also in the following cases :
\begin{enumerate}
\item
  The particle or the string in the pp-wave background in Brinkmann
  coordinates 
  that is described by the metric
  \begin{equation}
    d s^2_B = - 2 du\, d v
    + \sum_{I=1}^{D-2} A_I (A_I - 1) ( x^I)^2 \frac{1}{u^2}  d u^2
    + \sum_{I=1}^{D-2} ( d x^I)^2
    .
  \end{equation}
  Notice however that a purely gravitational
  string background, i.e. with trivial dilaton and Kalb-Ramond,
  must be a Ricci flat background
  so we need to impose $\sum _I A_I (A_I - 1) = 0$ if we want a
  consistent model propagating in this background.
  The particle action in light-cone gauge $u=\tau$ reads
  \begin{equation}
    S_{LC}
    =
    \int d \tau\,
    \left[
    \frac{-1}{e} \dot v
    +
    \sum_{I=1}^{D-2}
    \left(
    \frac{-1}{e} (\dot x^I)^2
    +
    \frac{-1}{e} \frac{A_I (A_I - 1)}{\tau^2} (x^I)^2
    \right) 
    \right].
  \end{equation}
  Since $e$ is constant on shell, any $x^I$ has the action
  \eqref{eq:L_Brinkmann_time_dep_harm_osc} with $\omega^2=0$.
  The case with $\omega^2\ne0$ is recovered when string is considered.
  In facts the previous $x^I$ are the string zero modes and
  the string non zero modes $x^I_n$ have $\omega^2 \propto n^2$.
  
\item
  The modes of the scalar field in de Sitter universe in conformal
  time.
  If we consider the FLRW metric
  \begin{equation}
    d s^2
    =
    d t^2 - a^2(t) \sum_{i=1}^{D-1} ( d x^i)^2
    =
    a^2(\eta) \left(d \eta ^2 - \sum_{i=1}^{D-1} ( d x^i)^2 \right)
    ,
  \end{equation}
  with $d \eta = \frac{1}{a(t)}d t$.
  For de Sitter we have $a_{d S}(t)=e^{H t}$ so that
  $a_{d S}(\eta)= - \frac{1}{H \eta}$
  with $-\infty< \eta< 0^-$.
  The real scalar action is then
  \begin{align}
    S_{FLRW}
    &=
    \int d t\, d^{D-1} x\, a(t)^D\,
    \left[
      \oh  (\dot \phi)^2 - \oh a(t)^{-2} (\partial_i \phi)^2
      - \oh m^2 \phi^2
      \right]
    \nonumber\\
    =&
    \int d \eta\, d^{D-1} x
    \Biggl[
      \oh  (\dot \chi)^2 - \oh  (\partial_i \chi)^2
      \nonumber\\
      &
      - \oh \left( m^2 a^2
      - \frac{D-2}{2} \frac{a''(\eta)}{a(\eta)}
      - \frac{(D-2)(D-4)}{4} \left(\frac{a'(\eta)}{a(\eta)} \right)^2      
      \right)
      \chi^2
      \Biggr]
    ,
  \end{align}
  where we defined $\phi(t,x) = a^{1- \frac{D}{2}}(\eta)
  \chi(\eta,x^i)$ and $a'(\eta) = \frac{d a(\eta)}{d \eta}$.
  Performing the Fourier transform \wrt{} to the space coordinates we get
  \begin{align}
    S_{FLRW}
    &=
     \int d \eta\, d^{D-1} k
    \Biggl[ 
      \oh  |\tilde \chi'(\eta,k)|^2
      \nonumber\\
      &
   - \oh \left( k_i^2 + m^2 a^2
      - \frac{D-2}{2} \frac{a''(\eta)}{a(\eta)}
      - \frac{(D-2)(D-4)}{4} \left(\frac{a'(\eta)}{a(\eta)} \right)^2      
      \right)
      |\tilde \chi(\eta,k)|
      \Biggr]
    ,
    \end{align}
  which in de Sitter space becomes
  \begin{align}
    S_{d S}
    &=
     \int d \eta\, d^{D-1} k
    \left[ 
      \oh  |\tilde \chi'(\eta,k)|^2
      - \oh \left( k_i^2 + \frac{m^2}{H^2} \frac{1}{\eta^2}
      - \frac{D(D-2)}{4} \frac{1}{\eta^2}
      \right)
      |\tilde \chi(\eta,k)|
      \right]
    ,
    \end{align}
  which shows that the modes again have
  action  \eqref{eq:L_Brinkmann_time_dep_harm_osc} but with $\eta<0$
  so he model we consider  is a kind of cyclic de Sitter.
\item
  The particle in Vaidya metric with linear mass.
  
\end{enumerate}

\subsection{Failure of the perturbative expansion of the evolution
  operator in the interaction picture}
Let us now consider the Hamiltonian corresponding to
\eqref{eq:L_Brinkmann_time_dep_harm_osc} as the sum of the usual
harmonic oscillator and a quadratic time dependent interaction term.
The splitting we perform between the
unperturbed Hamiltonian and the perturbative part is somewhat artificial
but it is chosen in order to get the simplest example as possible and
then discuss the issues in the simplest context.

Explicitly in Schroedinger picture we have
\begin{align}
  H_S(t) =& H_{S 0}(t) + H_{S 1}(t)
  \nonumber\\
  H_{S 0}(t) = \frac{p_S^2}{2} + \oh \omega^2 x_S^2,
  ~~&~~
  H_{S 1}(t) = \frac{k}{t^2} x_S^2
  .
\end{align}
Obviously the perturbation Hamiltonian is dominant for small $t$ and
therefore one can expect that perturbation theory be asymptotic as it
happens in Stark effect.
However we find a complete breakdown of perturbation theory and not an
asymptotic series.

The interaction picture is obtained from Schroedinger equation
\begin{equation}
  i \frac{\partial}{\partial t} |\psi_S(t,t_0)\rangle
  = H_S(t) |\psi_S(t,t_0)\rangle
  ,
\end{equation}
by defining
\begin{equation}
  |\psi_I(t,t_0)\rangle
  = U_{0 S}(t_0, t) |\psi_S(t,t_0)\rangle
  ,~~~~
  U_{0 S}(t_0, t)= T e^{-i \int^{t_0}_t d t' H_{0 S}(t')} 
  ,
\end{equation}
where $U_{0 S}$ is the evolution operator for the ``free'' Hamiltonian
$H_{0 S}$.
The new state $|\psi_I(t,t_0)\rangle$ then evolves as
\begin{align}
  i \frac{\partial}{\partial t} |\psi_I(t,t_0)\rangle
  =&
  H_I(t, t_0) |\psi_I(t,t_0)\rangle
  ,
  \nonumber\\
  H_I(t, t_0)
  =&
  U_{0 S}(t_0, t) H_{1 S} (t) U_{0 S}(t, t_0)
  .
\end{align}
The Schroedinger equation in interaction picture has then formal
and perturbative solution
\begin{equation}
  |\psi_I(t,t_0)\rangle
  =
  T e^{-i \int^{t}_{t_0} d t' H_{I}(t', t_0)} 
  |\psi_I(t_0,t_0)\rangle
  =
  \left(
  1
  -i \int^{t}_{t_0} d t' H_{I}(t', t_0)
  +\dots
  \right)
  |\psi_I(t_0,t_0)\rangle
  .
\end{equation}
If we apply this formalism to our specific case we obtain the
interaction Hamiltonian
\begin{equation}
  H_{I}(t,t_0)
  =
  - \frac{k}{4 \omega t^2}
  \left(
  e^{2 i \omega (t-t_0)} a_S^{\dagger 2}
  +
  e^{-2 i \omega (t-t_0)} a_S^{ 2}
  - a_S^{\dagger} a_S
  - a_S a_S^{\dagger}
  \right)
  ,
\end{equation}
where we have as usual 
\begin{equation}
  a_S = \frac{p_S -i \omega x_S}{\sqrt{2 \omega}}
  ,~~~~
  [a^\dagger_S, a_S]=1
  ,~~~~
  U_{0 S}(t, t_0) = e^{ -i \omega (a_S^{\dagger} a_S + \oh )(t-t_0)}
.
\end{equation}
We can then build a basis for the Hilbert space
$\{|n\rangle\}_{n\in\N}$ as
\begin{equation}
  a_S |0\rangle=0
  ,~~~~
  |n\rangle = \frac{a^{\dagger n}_S}{\sqrt{n!}} |0\rangle
.
  \end{equation}

It is then immediate to see that the first order in perturbative
expansion for the evolution operator from a negative $t_0<0$ time to a
positive time $t_1>0$ is infinite.
Explicitly,
if we evolve perturbatively from $|\psi_I(t_0,t_0)\rangle = |n\rangle$ to
$|\psi_I(t_1,t_0)\rangle$
and we try to expand $|\psi_I(t_1,t_0)\rangle$ on the basis
$\{|m\rangle\}$
we have
\begin{align}
  \langle m | \int^{t_1}_{t_0} d t' H_{I}(t', t_0) | n \rangle
  &=
  -
  \frac{k}{4 \omega}
  \delta_{m, n} (2 m +1)
  \int^{t_1}_{t_0} d t \frac{1}{t^2}
  \nonumber\\
  &-
  \frac{k}{4 \omega}
  \delta_{m, n+2} \sqrt{m (m-1)}
  \int^{t_1}_{t_0} d t \frac{e^{2 i \omega (t-t_0)}}{t^2}
  \nonumber\\
  &-
  \frac{k}{4 \omega}
  \delta_{m, n-2} \sqrt{(m+2) (m+1)}
  \int^{t_1}_{t_0} d t \frac{e^{-2 i \omega (t-t_0)}}{t^2}
  .
\end{align}
This shows that not only the amplitude is divergent but that we cannot
expand $|\psi_I(t_1,t_0)\rangle$ on the Hilbert basis moreover the divergence
cannot be reabsorbed into a c-number shift of the Hamiltonian since
all coefficients depend on the states.
For later use we notice that to this order of perturbation we have
\begin{equation}
  \langle m | \int^{t_1}_{t_0} d t' H_{I}(t', t_0) | n \rangle
  =
  \int^{t_1}_{t_0}
  \langle m_S(t', t_0) | H_{1 S}(t') | n_S(t', t_0) \rangle
  ,
\end{equation}
i.e. we can actually use the Schroedinger states and Hamiltonian
without actually computing the corresponding objects in the
interaction picture.

\subsection{The complete theory is well defined: the $H_B$ case}
Given the previous failure of the perturbative expansion one can
wonder whether the theory exists across the singularity.
The answer as we show is affirmative.
The same problem has been considered before in
\cite{deVega:1990kk,deVega:1990ke,deVega:1990gq,deVega:1991nm,Jofre:1993hd,Tolley:2003nx,Craps:2008bv,Madhu:2009jh,Narayan:2009pu,Narayan:2010rm} but our
point of view is slightly different since this is not the final
research target of
this paper but we want anyhow to show that we can traverse the
singularity and then use this solution for the interacting models.

Even if we are actually interested in adding quartic and higher
interactions to $L_R$ we will perform the analysis for $L_B$
since it looks more familiar
and then map it to $L_R$ using a time dependent unitary transformation.

The time dependent harmonic oscillator
\begin{equation}
  i \partial_t \psi(x,t)
  = -\oh \partial_x^2 \psi(x,t)
  + \oh \left( \omega^2 + \frac{k}{t^2} \right) \psi(x,t)
,
\end{equation}
can be solved exactly using
complex classical solutions with a well defined normalization.
We review the derivation for completeness in
appendix \ref{app:time_dep_harm_osc} where we give also more details
which are not relevant for the present discussion.
The main result is then that the generating function of a possible
complete set\footnote{Different sets are associated with different
instantaneous vacua.} of wave functions is
\begin{align}
    \sum_{n=0}^\infty \frac{z^n}{\sqrt{n!}} \psi_{ n\{\tz\}}(x, t, \tz)
  =&
     \sqrt[4]{\frac{1}{2 \pi}}
     \frac{1}{\sqrt{\cX(t)}}
     e^{i \frac{1}{2} \frac{ \dot \cX(t) }{ \cX(t)} x^2
     +
     \frac{1}{\cX(t)} x z
     -
     \oh \frac{\cX^*(t)}{\cX(t)} z^2
     }
     ,
\label{eq:generating_function_harm_osc}
\end{align}
where we have introduced the complex classical solution $\cX(t)$ and
its normalization condition
\begin{align}
  \ddot \cX(t) + \Omega^2(t) \cX(t)
  &= 0
    ,
  \nonumber\\
  \cX^* \dot \cX - \cX \dot \cX^*
  &=
    i 
  .
    \label{eq:Kr_cX_eqs}
\end{align}
We can now solve perturbatively the classical equations of motion
around $t=0$.

An issue which arises is the continuation across the singularity but
the normalization condition required for the quantum model and
``continuity'' fix it (see also \cite{Craps:2007iu} for the case $A=\oh$).

Let us start considering the asymptotic behavior for
$t\rightarrow 0^+$ as $\cX \sim t^a$ with $t>0$.
It is immediate to find the equation
\begin{equation}
a^2-a +k = 0
\Longleftrightarrow
a\in\{A, 1-A\}
,
\end{equation}
so that the leading behavior is
\begin{equation}
\cX(t) = c_0 (\omega t)^A (1+ O(t^2) ) + c_1 (\omega t)^{1-A} (1+ O(t^2) )
,~~~~ t>0
.
\label{eq:asymp_cX_positive_t}
\end{equation}
The normalization condition then implies
\begin{equation}
-(2 A - 1) \omega |c_1|^2 \Im\left( \frac{c_0}{c_1} \right) = - \oh
.
\label{eq:c0_c1_normalization}
\end{equation}
Let us consider the case $A>\oh> 1-A$ since $A<\oh<1-A$ is obtained by swapping
$A \leftrightarrow 1-A$.
Then the previous condition implies that the wave functions are
normalizable since ($t>0$)
\begin{align}
\psi_0(x,t)
\sim&
\frac{1}{\sqrt{|\omega t|^{1-A}}}
e^{ i \oh \left( \frac{1-A}{t}
+ (2 A-1) \omega \frac{c_0}{c_1} |\omega t|^{2(A-1)} \right) x^2
}
\nonumber\\
&\Longleftrightarrow
|\psi_0(x,t)|^2
\sim
\frac{1}{|\omega t|^{1-A}}
e^{ 
- (2 A-1) \omega \Im \left(\frac{c_0}{c_1}\right) |\omega t|^{2(A-1)} x^2
}
.
\label{eq:leading_behavior_time_dep_harm_osc_psi0}
\end{align}
As discussed in appendix around eq. \eqref{app:eq:time_dep_harm_osc_normalizability} this is not by chance: the normalization
condition on $\cX$ always implies the normalizability of the wave
functions.

Let us exam the solution for $t<0$. One would be tempted to write
exactly  the same equation \ref{eq:asymp_cX_positive_t} with the
substitution $t \rightarrow -t$.
However this would lead to a different normalization condition. The
difference being an overall sign in the left hand side of the
normalization equation, i.e. $+\oh$ in stead of $-\oh$.
Therefore the proper asymptotic behavior valid for all $t$ is 
either
\begin{equation*}
\cX(t) = c_0 |\omega t|^A (1+ O(t^2) ) + c_1 \omega t |\omega t|^{-A} (1+ O(t^2) )
,
\end{equation*}
or
\begin{equation}
\cX(t) = c_0 \omega t |\omega t|^{A-1} (1+ O(t^2) ) + c_1 |\omega t|^{1-A} (1+ O(t^2) )
.
\label{eq:asymp_cX}
\end{equation}
Since this is a classical solution we may expect that the trajectory
is continuous then for $A>1$ comparing $t |t|^{-A}$ and $|t|^{1-A}$ we
realize that only the latter is continuous.
Hence the true solution is \eqref{eq:asymp_cX}.
Because of this the previous expression for the wave
function \eqref{eq:leading_behavior_time_dep_harm_osc_psi0}
where we took care of distinguish between $t$ and $|t|$ is
valid for all $t$ values.

As discussed in appendix \ref{app:complex_classical_solutions}
 the previous choice can also be obtained
regularizing the time dependent pulsation $\Omega^2(t)=\omega^2+ \frac{k}{t^2}$.

It is also possible and instructive to use the WKB expansion.
We write $\psi(x,t) = e^{i S(x,t) }$ so that we have to solve the equation
\begin{equation}
   \partial_t S(x,t)
  +\oh (\partial_x S(x,t))^2
  + \oh \left( \omega^2 + \frac{k}{t^2} \right)
  - i \oh \partial_x^2 S(x,t)
  = 0
  .
\end{equation}
This is done in appendix \ref{app:WKB_analysis_time_dep_ham_osc}.

\begin{figure}[hbt]
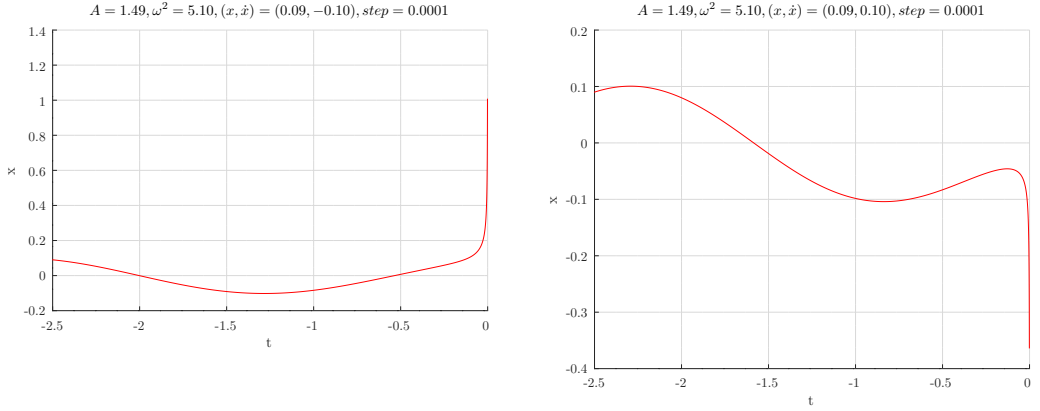

   \begin{subfigure}[b]{0.45\textwidth}
     \scalebox{.5}{
\input{Brinkmann__n=2__A_1.49__w2_5.10____x,v_____0.09,-0.10______t_i,t_f__=__-2.500000,-0.001000____step_0.000100.tex}}
    \caption{$\dot x=-0.10$ has $x(0)=+\infty$}
\label{fig:A_gt_1a}
   \end{subfigure}
  \hspace{0.1\textwidth}
   \begin{subfigure}[b]{0.45\textwidth}
     \scalebox{.5}{
       \input{Brinkmann__n=2__A_1.49__w2_5.10____x,v_____0.09,0.10______t_i,t_f__=__-2.500000,-0.001000____step_0.000100.tex}}
    \caption{
    $\dot x=+0.10$ has $x(0)=-\infty$}
\label{fig:A_gt_1b}
   \end{subfigure}
   \caption{ Classical motion for $L_B$ with $A>1$ has two possible
   asymptotic behaviors} 
\label{fig:Classical_motion_time_dep_ho_A_gt_1}
\end{figure}

Notice that \ref{eq:leading_behavior_time_dep_harm_osc_psi0}
has two completely different behaviors as $t \rightarrow 0$.
\begin{equation}
|\psi_0(x,t)|^2
\sim_{t \rightarrow 0}
\left\{
\begin{array}{l c}
0 & A>1
\\
\infty & A<1
\end{array}
\right.
.
\end{equation}
This can be understood considering the classical trajectory which
behaves as $x\sim |t|^{min(A, 1-A)}$. For $A>1$ it diverges but the
direction depends on the initial $\dot x$ which quantum mechanically
cannot be fixed therefore the quantum state is spread over all the
possible values of $x$. This is shown in figures \ref{fig:A_gt_1a} and
\ref{fig:A_gt_1b}.
Notice that the classical trajectory (not the complex one used in
computing the quantum wave function) is not well defined through $t=0$
since we can require the continuity of the trajectory but it is
difficult if not impossible to relate the velocity
before and after the singularity.
On the contrary the quantum theory is well defined since we can find a
well defined basis of wave functions. 

\begin{figure}[hbt]
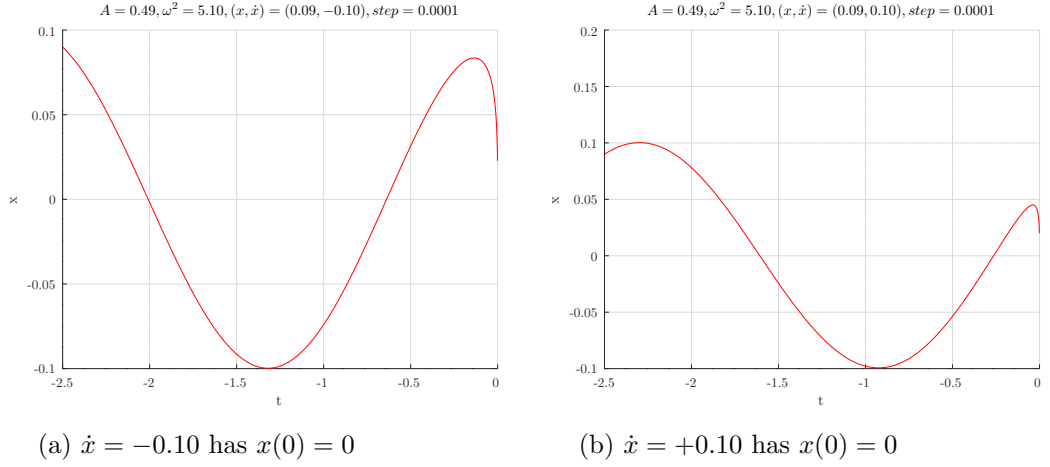

   \begin{subfigure}[t]{0.45\textwidth}
     \scalebox{.5}{
       \input{Brinkmann__n=2__A_0.49__w2_5.10____x,v_____0.09,-0.10______t_i,t_f__=__-2.500000,-0.001000____step_0.000100.tex}}
    \caption{
    $\dot x=-0.10$ has $x(0)=0$}
\label{fig:A_lt_1a}
   \end{subfigure}
  \hspace{0.1\textwidth}
   \begin{subfigure}[t]{0.45\textwidth}
     \scalebox{.5}{
       \input{Brinkmann__n=2__A_0.49__w2_5.10____x,v_____0.09,0.10______t_i,t_f__=__-2.500000,-0.001000____step_0.000100.tex}}
    \caption{
    $\dot x=+0.10$ has $x(0)=0$}
\label{fig:A_lt_1b}
   \end{subfigure}
   \caption{ Classical motion for $L_B$ with $A<1$ has only one possible
   asymptotic behavior} 
\label{fig:Classical_motion_time_dep_ho_A_lt_1}
\end{figure}

Differently for $A<1$  the classical  solution has a fixed point
$x(0)=0$
and therefore the wave function is a $\delta(x)$.
This is shown in figures \ref{fig:A_lt_1a} and \ref{fig:A_lt_1b}. 

Finally notice that the wildly oscillating phase
in \ref{eq:leading_behavior_time_dep_harm_osc_psi0} is not an issue
as hypothesized in \cite{Craps:2008bv,Madhu:2009jh,Narayan:2009pu,Narayan:2010rm},
on the contrary as shown in \cite{Arduino:2020axy} it is a virtue
since it helps the
convergence of the integrals in the distributional sense (see also \cite{vindas2007distributional}).

\subsection{Relation between $L_R$ and $L_B$ non interacting models}
While at the classical level the two models are related as described
before by a simple change of coordinates and a boundary term, at the
quantum level we have
\begin{equation}
  \psi_B(x, t)
  =
  |t|^{-\oh A}
  e^{ i \oh A \frac{x^2}{t}}
  \psi_R(y =   |t|^{-A} x, t)
  .
  \label{eq:psiB_psiR_non_interacting}
\end{equation}
This can be obtained in two different ways.
Both start from the Hamiltonians
\begin{align}
  H_R
  =&
  \frac{p_y^2}{2 |t|^{2 A} } + \oh \omega^2 |t|^{2 A} y^2
  \nonumber\\
  H_B
  =&
  \frac{p^2}{2  } + \oh \left(\omega^2 + \frac{k}{t^2} \right)  x^2
.
\end{align}
The first method is a sequence of transformations on the Schroedinger
equation.
We first change variables from
\begin{align}
\left\{
\begin{array}{c}
x= |t|^{A} y
\\
\tilde t = t
\end{array}
\right.
~~\Rightarrow~~
\left\{
\begin{array}{c}
\frac{\partial}{\partial t}
=
\frac{\partial}{\partial  \tilde t}
+ \frac{A x}{\tilde t} \frac{\partial}{\partial x} 
\\
\frac{\partial}{\partial y}
=
|\tilde t|^A
\frac{\partial}{\partial x}
\end{array}
\right.
.
\end{align}
Then the $H_R$ Schroedinger equation becomes
\begin{equation}
i \frac{\partial}{\partial \tilde t} \hat \psi(x, \tilde t)
=
\left(
-\oh \partial_x^2 + \oh \omega^2 x^2
- i \frac{A x}{\tilde t} \partial_x
\right)
\hat \psi(x, \tilde t)
,
\end{equation}  
with $\psi_R(y, t) = \hat \psi(x, \tilde t)$.
However this equation is not a Schrodinger equation since the would be
Hamiltonian is not Hermitian because of the term
$- i \frac{A x}{\tilde t} \partial_x$.
To get an Hermitian Hamiltonian we redefine
$\hat \psi(x, \tilde t)= |\tilde t|^{\oh A}  \psi_I(x, \tilde t)$.
Notice that the factor $|\tilde t|^{\oh A}$ is the factor one could expect
from the measure due to the change $x= |t|^{A} y$.
We get then the intermediate Schroedinger equation
\begin{equation}
i \frac{\partial}{\partial \tilde t} \psi_I(x, \tilde t)
=
\left[
\oh \left( -i \partial_x + \frac{A x}{\tilde t}\right)^2
+ \oh \left( \omega^2 - \frac{A^2}{2 \tilde t^2} \right) x^2
\right]
\psi_I(x, \tilde t)
,
\end{equation}  
with
$\psi_R(y, t) = |\tilde t|^{\oh A}  \psi_I(x, \tilde t)$.
Finally we make a further redefinition as
$\psi_I(x, \tilde t) =  e^{-i \oh \frac{A^2}{\tilde t^2} x^2} \psi_B(x, \tilde t)$
in order to have a canonical kinetic term.
We finally get the desired result
\begin{equation}
i \frac{\partial}{\partial \tilde t} \psi_B(x, \tilde t)
=
\left[
- \oh \partial_x^2
+ \oh \left( \omega^2 + \frac{A-A^2}{2 \tilde t^2} \right) x^2
\right]
\psi_B(x, \tilde t)
,
\end{equation}  
where the relation between $\psi_R$ and $\psi_B$ is the one given
above in \eqref{eq:psiB_psiR_non_interacting}.

The second method is operatorial.
The first step is to use a unitary transformation which implements
\begin{align}
\left\{
\begin{array}{c l}
x= |t|^{A} y &= U_{R\rightarrow I}^\dagger\, y\, U_{R\rightarrow I} 
\\
p = \frac{p_y}{|t|^{A}} &= U_{R\rightarrow I}^\dagger\, p_y\, U_{R\rightarrow I} 
\end{array}
\right.
~~
\Rightarrow
~~
U_{R\rightarrow I}
=
e^{ i \ln(|t|^{A})\, \oh\{y, p_y\} }
=
|t|^{\oh A} |t|^{i A y p_y}
.
\end{align}
We then get the intermediate Hamiltonian\footnote{
The term $i \dot  U^\dagger\, U$ is
obtained from the Schroedinger equation as follows.
Set $|\psi_I(t)\rangle = U^\dagger(t) |\psi_R(t)\rangle$
then from $i \partial_t |\psi_R(t)\rangle = H_R |\psi_R(t)\rangle$
we get $i \partial_t |\psi_I(t)\rangle = H_I |\psi_I(t)\rangle$
with
$H_I = U^\dagger\, H_R\, U + i \dot  U^\dagger\,  U $.
}
\begin{align}
H_I
=&
U_{R\rightarrow I}^\dagger\, H_R\, U_{R\rightarrow I} 
+
i \dot  U_{R\rightarrow I}^\dagger\,  U_{R\rightarrow I} 
\nonumber\\
&=
  \frac{1}{2  } \left( p_y - \frac{A}{t} y \right)^2
  + \oh \left(\omega^2 - \frac{A^2}{t^2} \right)  y^2
.
\end{align}     
With a further unitary transformation
\begin{align}
\left\{
\begin{array}{c l}
 y &= U_{I\rightarrow B}^\dagger\, y\, U_{I\rightarrow B} 
\\
p_y - \frac{A}{t} y &= U_{I\rightarrow B}^\dagger\, p_y\, U_{I\rightarrow B} 
\end{array}
\right.
~~
\Rightarrow
~~
U_{I\rightarrow B}
=
e^{ -i \oh \frac{A}{t} y^2}
,
\end{align}
used to make the kinetic term canonical
we finally get the desired result. Explicitly
\begin{align}
H_B
=&
U_{I\rightarrow B}^\dagger\, H_I\, U_{I\rightarrow B} 
+
i \dot  U_{I\rightarrow B}^\dagger\,  U_{I\rightarrow B} 
\nonumber\\
&=
  \frac{1}{2  } p_y^2
  + \oh \left(\omega^2 + \frac{A-A^2}{t^2} \right)  y^2
,
\end{align}     
so that
\begin{equation}
|\psi_B(t)\rangle
=
U_{I\rightarrow B}^\dagger
U_{R\rightarrow I}^\dagger
|\psi_R(t)\rangle
,
\label{eq:B_to_R_unitary_transformation_on_psi}
\end{equation}
which again reproduces \eqref{eq:psiB_psiR_non_interacting}.

\subsection{Explicit mapping of the quantum $H_B$ solutions to $H_R$ solutions}
Using the explicit mapping in \eqref{eq:psiB_psiR_non_interacting} we
can write the generating function for a complete set of solutions for
$H_R$ as
\begin{align}
    \sum_{n=0}^\infty \frac{z^n}{\sqrt{n!}} \psi_{R\, n\{\tz\}}(y, t, \tz)
  =&
     \sqrt[4]{\frac{1}{2 \pi}}
     \frac{1}{\sqrt{\cX_R(t)}}
     e^{i \frac{1}{2} \frac{ \dot \cX_R(t) }{ \cX_R(t)} x^2
     +
     \frac{1}{\cX_R(t)} x z
     -
     \oh \frac{\cX_R^*(t)}{\cX_R(t)} z^2
     }
     ,
\label{eq:Rosen_generating_function_harm_osc}
\end{align}
where we have introduced the complex classical solution
$\cX_R(t)= |t|^{-A} \cX(t)$ in analogy to $y= |t|^{-A} x$.
Its \eom{} and normalization condition follow from the $\cX$ ones
and read
\begin{align}
  |t|^{-2A} \frac{d}{d t} \left( |t|^{2A} \dot\cX_R(t) \right) + \omega^2 \cX_R(t)
  &= 0
    ,
  \nonumber\\
  \cX_R^* \dot \cX_R - \cX_R \dot \cX_R^*
  &=
    i |t|^{-2A}
    .
      \label{eq:Rosen_Kr_cX_eqs}
\end{align}

In particular the ``ground state'' behaves as
\begin{align}
\psi_{R\,0}(y,t)
\sim&
|t|^{\oh(A-1)}
e^{ i \oh \left( (1-A) sgn(t) |t|^{2A-1}
+ (2 A-1) \frac{c_0}{c_1} \omega^{2A-1} |t|^{2(2 A-1)} \right) y^2
}
\nonumber\\
&\Longleftrightarrow
|\psi_{R\,0}(y,t)|^2
\sim
|t|^{2A-1}
e^{ 
- (2 A-1) \omega \Im \left(\frac{c_0}{c_1}\right) \omega^{2A-1}
|t|^{2(2 A-1)} y^2
}
.
\label{eq:Rosen_leading_behavior_time_dep_harm_osc_psi0}
\end{align}
The wave functions always vanish for $t\rightarrow 0$ while still being
normalizable because the classical particle is diffused on the entire
$y$ axis since $y\sim |t|^{-2 A}$.
This         diverges but the
direction depends on the initial $\dot y$ which quantum mechanically
cannot be fixed.
\section{Interacting quantum and classical mechanical models}
\label{sect:interacting_qm_model}
We can now pass to exam what happens when we add interactions to the
Kasner metrics.
The corresponding quantum mechanical models are
\begin{equation}
  L_R = |t|^{2 A} \left(
  \oh \dot y^2 -\oh \omega^2 y^2
  - \frac{g}{n} y^n
  \right)
  ,~~~~
  g>0
  ,~~
  n\in\{4,6,\dots\}
  ,
\end{equation}
which become in $x$ coordinate
\begin{equation}
L_B = \oh \dot x^2
- \oh \left(\omega^2 + \frac{k}{t^2} \right) x^2
- \frac{g}{n} \frac{1}{ |t|^{A (n-2)} } x^n
.
\end{equation}
These models show a strange time dependence in the interaction term
which can be explained by noticing that the change from $y$ to $x$
in quantum mechanical models cannot be implemented on the metric.

The B models suggest that the
interaction is dominant for small $\omega t$.
This is not evident in R models and it is not always true.

Using the results from the previous section on the behavior of the
wave function at $t=0$ we can now see that the perturbative
expansion of the evolution matrix in interaction picture does not
exist.
Explicitly for B models (since they are unitarily equivalent to R models
with as in \eqref{eq:B_to_R_unitary_transformation_on_psi})
\begin{align}
  \int d t' \langle \psi_B(t') | H_{B\, S 1}(t') |\psi_B(t')\rangle
  \sim&
  \int d t'  \frac{1}{ |t|^{A (n-2)} }
  \int d x\, x^n\, |t'|^{-\alpha} e^{ - |t'|^{-2\alpha} x^2 }
  \nonumber\\
  \sim&
  \int d t'  \frac{1}{ |t'|^{A (n-2)} }
  \left(\frac{1}{|t'|^{-2\alpha}}\right)^\frac{n}{2}
  ,
\end{align}
which has an unavoidable divergence for $A>1$ and $-\alpha=A-1>0$.
More precisely the integral is divergent for $2A> \frac{n+1}{n-1}$.
Anticipating
the results (discussed below eq. \eqref{eq:interacting_model_alpha_beta}
for the classical case and around
eq. \eqref{eq:wave_functions_interacting_model_alpha_positive} for the
quantum case)
this means that when the behavior is dominated by the
interaction, i.e. $2 A > \frac{n+2}{n-2}$ the integral is divergent.
This integral may also be divergent when the theory is dominated by the
kinetic term,
i.e $\frac{n+2}{n}< 2A < \frac{n+2}{n-2}$ (see
eq. \eqref{eq:classical_a_lt_0_possibilities} and
eq. \eqref{eq:quantum_wave_a_lt_0}).

\subsection{The classical motion}
The classical \eom{} for the R models reads
\begin{equation}
|t|^{-2 A} \frac{d}{d t} \left( |t|^{2 A} \frac{d y}{d t} \right)
+ \omega^2 y + g y^{n-1}
=0
.
\end{equation}
This equation is very close to the Emden-Fowler equation
\begin{equation}
\frac{d}{d t} \left( t^{\mu} \frac{d y}{d t} \right)
+ t^\nu y^{m}
=0
.
\end{equation}
This equation is treated in \cite{bellman2008stability}
with the result that (with the appropriate range of the parameters
$\mu, \nu$ which can be easily obtained from our treatment) the solution
exhibits an oscillating behavior with maxima and minima diverging with
a power law.  
Instead of the analysis presented there
we introduce a different approach which is simpler and clearer based on
the action.
We apply immediately this approach to the R models whose action is
\begin{equation}
S_R = \int_I d t\,  |t|^{2 A} \left(
  \oh \dot y^2 -\oh \omega^2 y^2
  - \frac{g}{n} y^n
  \right)
  ,
\end{equation}
where $I$ is the integration interval.
We look for a change of variables as
\begin{equation}
t = sgn(\tilde t ) |\tilde t|^\beta
,~~~~
y = |\tilde t|^\alpha z
,
\label{eq:interacting_change_of_variables}
\end{equation}
so that the kinetic term  and the interaction term $z^n$
have  coefficients independent of the new time $\tilde t$.
Explicitly we get
\begin{align}
S_R
&=
\int_{\tilde I} d \tilde t\,
\Bigl\{
\oh \frac{1}{\beta}
|\tilde t|^{(2A-1)\beta + 2\alpha+1}
\left( \frac{d z}{d \tilde t} - \frac{\alpha}{\tilde t} z \right)^2
\nonumber\\
&
\phantom{\int_{\tilde I} d \tilde t\,}
-
\oh \beta \omega^2 |\tilde t|^{(2A+1)\beta + 2\alpha-1}
z^2
-
\beta \frac{g}{n} |\tilde t|^{(2A+1)\beta + n\alpha- 1}
z^n
\Bigr\}
,
\end{align}     
where $\tilde I$ is the image of the interval $I$.
We can now require a time independent kinetic and $z^n$ term imposing 
\begin{align}
(2A-1)\beta + 2\alpha+1=0
,~~~~
(2A+1)\beta + n\alpha- 1=0
,
\end{align}
which can be solved as
\begin{equation}
\alpha
=
\frac{4 A}{ 2(n-2) A - (n+2)}
,~~~~
\beta
=
-\frac{n+2}{ 2(n-2) A - (n+2)}
,
\label{eq:interacting_model_alpha_beta}
\end{equation}
and get
\begin{align}
S_R
&=
\int_{\tilde I} d \tilde t\,
\Bigl\{
\oh \frac{1}{\beta}
\left( \frac{d z}{d \tilde t} - \frac{\alpha}{\tilde t} z \right)^2
%
-
\beta \oh \omega^2 |\tilde t|^{(2-n)\alpha} 
z^2
-
\beta \frac{1}{n} g 
z^n
\Bigr\}
.
\label{eq:Action_z^n}
\end{align}     

The previous action can be recast in a more standard form by
integrating by part the term proportional to
$\frac{d z}{d \tilde t} z =  \oh \frac{d z^2}{d \tilde t}$ to get
\begin{align}
S_R
&=
\left.
+\oh \frac{\alpha}{\beta}
\frac{1 
}{\tilde t} z^2
\right|_{\tilde I}
\nonumber\\
&+
\int_{\tilde I} d \tilde t\,
\Bigl\{
\oh \frac{1}{\beta}
\left( \frac{d z}{d \tilde t} \right)^2
%
+
\Bigl[
\oh \frac{ \alpha \left( \alpha + 1 
\right)}{\beta}
-
\beta \oh \omega^2 \frac{1}{|\tilde t|^{(n-2)\alpha}}
\Bigr] z^2
%
-
\beta \frac{g}{n} 
z^n
\Bigr\}
.
\label{eq:Action_z^n_integrated_by_part}
\end{align}     

If $\alpha>0>\beta$ for $A>\oh+\frac{2}{n+2}$ the interval around the
singularity $t=0$
 $I=[-\epsilon_1, +\epsilon_2]$ is mapped into an interval
around $|\tilde t|=\infty$ as
$\tilde I
= [-\infty, -\frac{1}{\epsilon_1}] \cup
[\frac{1}{\epsilon_2}, +\infty]$
then the $z^2$ terms are subdominant since
$|\tilde t|^{(2A-1)\beta + 2\alpha+1}
=
\frac{1}{\tilde t^2}
$
and
$|\tilde t|^{(2A+1)\beta + 2\alpha-1}
=
\frac{1}{|\tilde t|^{(n-2) \alpha}}$.
Moreover the boundary term is finite.

Under the previous choice of $\alpha, \beta$ we can approximate 
the action $S_R$
for the $I$ around the singularity simply as\footnote{The fact
that $\beta<0$ is compensated by the orientation of $\tilde I$.}
\begin{align}
S_R
&\sim
\int_{\tilde I} d \tilde t\,
\Bigl\{
\oh \frac{1}{\beta}
\left( \frac{d z}{d \tilde t} \right)^2
-
\beta \frac{g}{n}
z^n
\Bigr\}
.
\end{align}     
Hence the trajectory $z(\tilde t)$ is simply oscillating with period
\begin{equation}
\oh P
=
\frac{1}{\sqrt{2 |\beta| E_z}}
\left( \frac{n E_z}{|\beta| g} \right)^{\frac{1}{n} }
\int_{-1}^{+1} d \zeta \frac{1}{\sqrt{1 - \zeta^n}}
,
\end{equation}  
where $E_z$ is the system energy.

\begin{figure}[hbt]
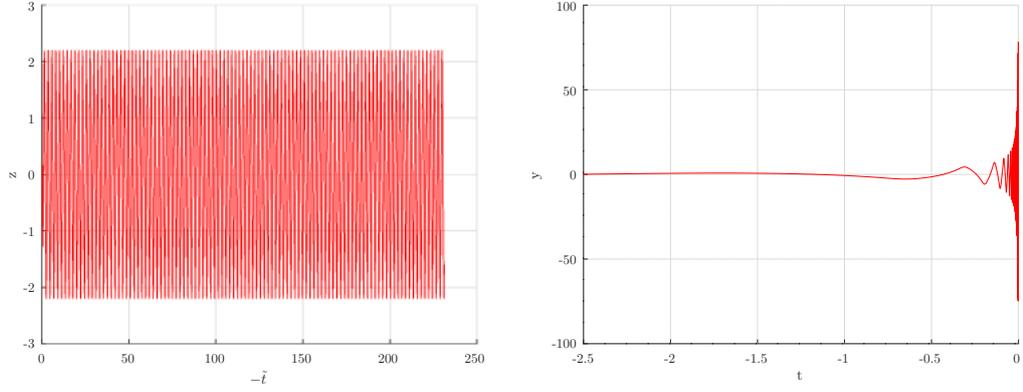

\begin{subfigure}[t]{0.45\textwidth}
     \scalebox{.5}{
       \input{Interacting__Redefined__n=2__A_1.49__w2_5.10____z,dot__z_____0.16,-12.16______g,n_____0.10,__8______t_i,t_f__=__-2.500000,-0.001000______tt_i,tt_f__=__0.485762,231.206479____step_0.00010.tex}}
    \caption{Motion in $z$ coordinate 
    and $-\tilde t$ time where the singularity is at $\tilde t=+\infty$}
\label{fig:z_tilte_t_n=8}
\end{subfigure}
\hspace{0.1\textwidth}
\begin{subfigure}[t]{0.45\textwidth}
     \scalebox{.5}{
       \input{Interacting__Rosen__n=2__A_1.49__w2_5.10____y,dot__y_____0.09,1.10__,____g,n_____0.10,__8______t_i,t_f__=__-2.500000,-0.001000____step_0.00010.tex}}
    \caption{The previous motion in $y$ coordinate (with remapped
       initial conditions) and $t$ time.}
\label{fig:y_t_n=8}
\end{subfigure}
   \caption{ Classical motion with $\alpha>0$
   }
\label{fig:Classical_motion_zn_alpha_gt_0}

\end{figure}

\begin{figure}[hbt]
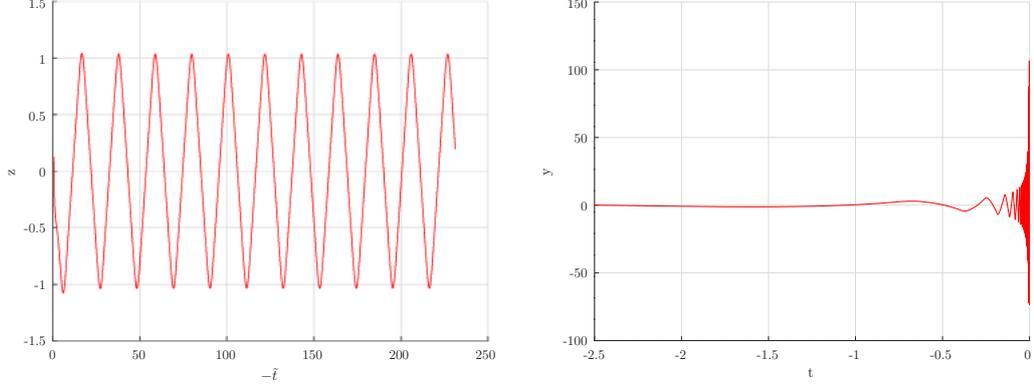

\begin{subfigure}[t]{0.45\textwidth}
     \scalebox{.5}{
       \input{Interacting__Redefined__n=2__A_1.49__w2_5.10____z,dot__z_____0.09,1.10______g,n_____0.10,__8______t_i,t_f__=__-2.500000,-0.001000______tt_i,tt_f__=__0.485762,231.206479____step_0.00010.tex}}
    \caption{Motion in $z$ coordinate 
    and $-\tilde t$ time where the singularity is at $\tilde t=+\infty$}
\label{fig:z_tilte_t_n=8_smoother}
\end{subfigure}
\hspace{0.1\textwidth}
\begin{subfigure}[t]{0.45\textwidth}
     \scalebox{.5}{
       \input{Interacting__Rosen__n=2__A_1.49__w2_5.10____y,dot__y_____0.05,-1.46__,____g,n_____0.10,__8______t_i,t_f__=__-2.500000,-0.001000____step_0.00010.tex}}
    \caption{The previous motion in $y$ coordinate (with remapped
       initial conditions) and $t$ time.}
\label{fig:y_t_n=8_smoother}
\end{subfigure}
   \caption{ Another classical motion with $\alpha>0$ with a smoother behavior
   }
\label{fig:Classical_motion_zn_alpha_gt_0_smoothe}
\end{figure}

Despite this nice feature the crossing of the singularity is not very
well defined at the classical level since $t=0^\pm$ is mapped to
$\tilde t= \pm \infty$ and there the particle is spread over the
interval $[-\left( \frac{n E_z}{|\beta| g} \right)^{\frac{1}{n} }
, \left( \frac{n E_z}{|\beta| g} \right)^{\frac{1}{n} }
]$ in $z$ coordinate
and it is not obvious how to match the position at 
$\tilde t= +\infty$ with the position at
$\tilde t= - \infty$.
This is shown in figures \ref{fig:z_tilte_t_n=8} and
\ref{fig:y_t_n=8} for $t\rightarrow 0^-$, i.e. for $\tilde
t\rightarrow -\infty$. 
And in a smoother case in \ref{fig:z_tilte_t_n=8_smoother} and \ref{fig:y_t_n=8_smoother}.

\begin{figure}[hbt]
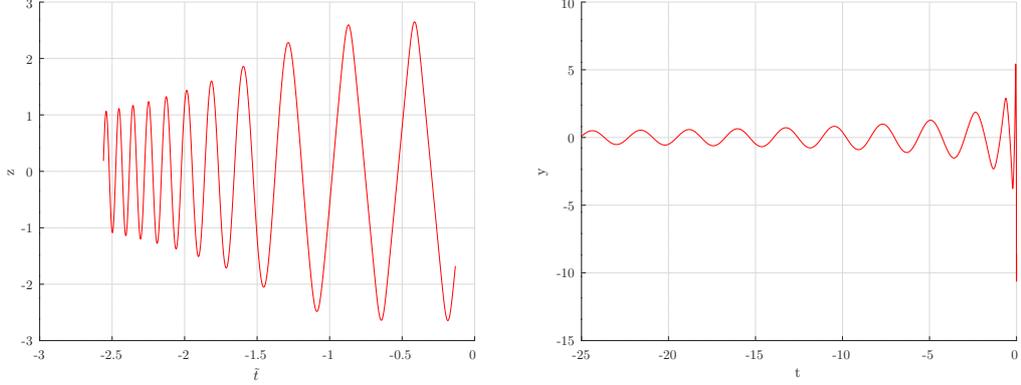

\begin{subfigure}[t]{0.45\textwidth}
     \scalebox{.5}{
       \input{Interacting__Redefined__n=2__A_0.59__w2_5.10____z,dot__z_____0.19,78.58______g,n_____0.10,__8______t_i,t_f__=__-25.000000,-0.001000______tt_i,tt_f__=__-2.559755,-0.133045____step_0.00010.tex}}
    \caption{Motion in $z$ coordinate with $\alpha<0$
       and $\tilde t$
       time where the singularity is at $\tilde t=0$.}
\label{fig:A_lt1_z_tilte_t_n=8}
\end{subfigure}
\hspace{0.1\textwidth}
\begin{subfigure}[t]{0.45\textwidth}
     \scalebox{.5}{
     \input{Interacting__Rosen__n=2__A_0.59__w2_5.10____y,dot__y_____0.09,1.10__,____g,n_____0.10,__8______t_i,t_f__=__-25.000000,-0.001000____step_0.00010.tex}}
    \caption{The previous motion (with remapped initial conditions) in $y$ coordinate and $t$ time.}
\label{fig:A_lt1_y_t_n=8}
\end{subfigure}
   \caption{ Classical motion with $\alpha<0$ with a too short
   integration range to show the expected behavior
   }
\label{fig:Classical_motion_zn_alpha_lt_0_wrong}
\end{figure}

\begin{figure}[hbt]
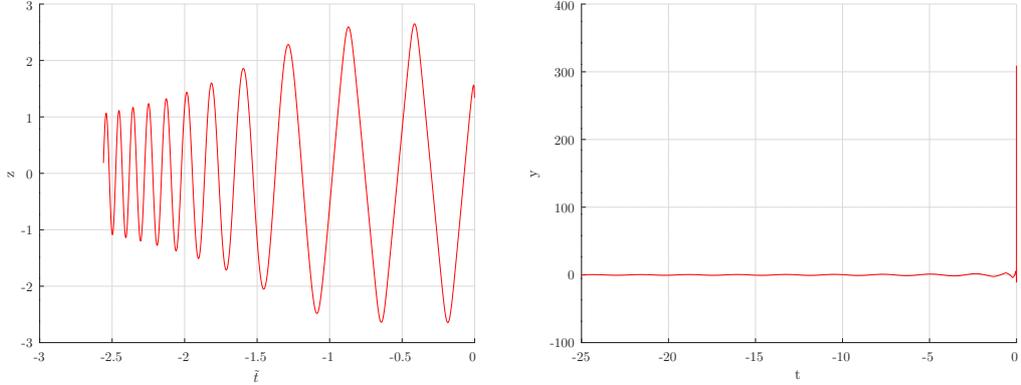

\begin{subfigure}[t]{0.45\textwidth}
     \scalebox{.5}{
       \input{Interacting__Redefined__n=2__A_0.59__w2_5.10____z,dot__z_____0.19,78.58______g,n_____0.10,__8______t_i,t_f__=__-25.000000,-0.000000______tt_i,tt_f__=__-2.559755,-0.001202____step_0.00010.tex}}
    \caption{Motion in $z$ coordinate 
    with the same
       parameters as in figure \ref{fig:A_lt1_z_tilte_t_n=8} but with
       an integration range which extends closer to the origin and shows
       the proper asymptotic.}
\label{fig:A_lt1_z_tilte_t_n=8_closer_0}
\end{subfigure}
\hspace{0.1\textwidth}
\begin{subfigure}[t]{0.45\textwidth}
     \scalebox{.5}{
     \input{Interacting__Rosen__n=2__A_0.59__w2_5.10____y,dot__y_____0.09,1.10__,____g,n_____0.10,__8______t_i,t_f__=__-25.000000,-0.000000____step_0.00010.tex}}
    \caption{The previous motion (with remapped initial conditions) in $y$ coordinate and $t$ time.}
\label{fig:A_lt1_y_t_n=8_closer_0}
\end{subfigure}
   \caption{ Classical motion with $\alpha<0$ with a proper
   integration range to show the expected behavior
   }
\label{fig:Classical_motion_zn_alpha_lt_0_right}
\end{figure}

For the case $\alpha<0<\beta$ the behavior of the classical motion
is dictated by 
\begin{align}
S_R
&\sim
\int_{-|\tilde \epsilon_1|}^{-|\tilde \epsilon_2|} d \tilde t\,
\frac{1}{\beta}
\Bigl\{
\oh 
\left( \frac{d z}{d \tilde t} \right)^2
+
\oh
\alpha(\alpha+1)
\frac{1}{|\tilde t|^2}
z^2
\Bigr\}
\nonumber\\
&\sim
\int_{-|\tilde \epsilon_1|}^{-|\tilde \epsilon_2|} d \tilde t\,
\frac{1}{\beta}
\Bigl\{
\oh 
\left( \frac{d z}{d \tilde t} \right)^2
+
\oh
\frac{4A ( 2 n A - (n+2) )}{ ( 2(n-2) A - (n+2) )^2}
\frac{1}{|\tilde t|^2}
z^2
\Bigr\}
,
\end{align}     
because the boundary term does not contribute to the \eom{} we find
again a time dependent harmonic oscillator as in
eq. \eqref{eq:L_Brinkmann_time_dep_harm_osc} but with $A_{eff}$ (where
$k_{eff}= - \alpha(1+\alpha) = A_{eff} (1-A_{eff})$, i.e. $A_{eff}=-\alpha$) which
is always real, explicitly
\begin{equation}
\left\{
\begin{array}{c c c}
k_{eff} > \frac{1}{4} & A_{eff}\in\C
& \mbox{not possible}
\\
0< k_{eff} < \frac{1}{4} & 0\le A_{eff}\le 1 
&  2 A < \frac{n+2}{n}
\\
k_{eff} < 0 & A_{eff}>1
& \frac{n+2}{n} < 2 A < \frac{n+2}{n-2}
\end{array}
\right.
,
\label{eq:classical_a_lt_0_possibilities}
\end{equation}
As usual numerics can be tricky and give the wrong impression: compare
the figures \ref{fig:A_lt1_z_tilte_t_n=8} and \ref{fig:A_lt1_y_t_n=8}
with the same solution extended closer to the origin given in figures
\ref{fig:A_lt1_z_tilte_t_n=8_closer_0} and \ref{fig:A_lt1_y_t_n=8_closer_0}.
Both for $t\rightarrow 0^-$, i.e for $\tilde t\rightarrow 0^-$

%
%

\subsection{The quantum interacting models exist}
We can now exam the question of what happens to the quantum model.
We treat only the wave function approach because it is more intuitive.

Despite the fact the classical motion is not very well defined the
quantum system seems to be perfectly fine and generically better
behaved than the non interacting one.
The last sentence means that we can write a normalizable wave function
which generically vanishes at $t=0$ but at slower rate that the non interacting,
i.e. quadratic R theory.
The adverb generically refers to the fact that there is a ``small'' range of
parameters where system behavior can be mapped to a time dependent
harmonic oscillator with unbounded potential.

Another point to stress is that we have found a possible continuation
through the singularity it may be that there are other possibilities
as in the free case \cite{Craps:2007iu}.

In order to show that we start with Schroedinger equation for R model
\begin{align}
i \partial_t \psi(y,t)
=
\left[
-\oh \frac{1}{|t|^{2A}} \partial_y^2
+ |t|^{2A} \left( \oh \omega^2 y^2 + \frac{g}{n} y^n \right)
\right] \psi(y,t)
,
\end{align}
and following the previous section on the classical motion
we perform the same change of variables as in the classic case \eqref{eq:interacting_change_of_variables}
\begin{align}
\left\{
\begin{array}{c}
\tilde t = sgn(t) |t|^{\frac{1}{\beta}}
\\
z= |t|^{-\frac{\alpha}{\beta}} y
\end{array}
\right.
~~\Rightarrow~~
\left\{
\begin{array}{c}
\frac{\partial}{\partial t}
=
\frac{|\tilde t|^{-\beta+1}}{\beta}
\left(
\frac{\partial}{\partial  \tilde t}
- \frac{\alpha z}{\tilde t} \frac{\partial}{\partial x} 
\right)
\\
\frac{\partial}{\partial y}
=
|\tilde t|^{-\alpha}
\frac{\partial}{\partial z}
\end{array}
\right.
,
\end{align}
along with setting $\psi(y,t)= |\tilde t|^{-\oh \alpha} \tilde \psi(z,\tilde t)$.
The choice of the $\tilde t$ power is made considering the invariance of the
probability density $|\psi(y,t)|^2 d y = |\tilde \psi(z,\tilde t)|^2 d
z$.
The Schroedinger equation then becomes
\begin{align}
i \frac{1}{\beta} \frac{\partial}{\partial \tilde t} \tilde \psi(z,\tilde t)
=&
-\oh
|\tilde t|^{-(2A-1)\beta -2\alpha-1}
\frac{\partial^2}{ \partial z^2 } \tilde \psi(z,\tilde t)
\nonumber\\
&+
\frac{g}{n} 
|\tilde t|^{(2A+1)\beta +n\alpha-1}
z^n \tilde \psi(z,\tilde t)
+
\oh \omega^2
|\tilde t|^{(2A+1)\beta +2\alpha-1}
z^2 \tilde \psi(z,\tilde t)
\nonumber\\
&
+i
\frac{\alpha}{2 \beta}
\frac{1}{\tilde t}
\left( z \frac{\partial}{\partial z} + \frac{\partial}{\partial z}
z\right) \tilde \psi(z,\tilde t)
.
\end{align}
If we require the kinetic and  $z^n$ terms
be time independent
we get exactly the same solution for $\alpha, \beta$ as in the
classical case \eqref{eq:interacting_model_alpha_beta} and the
Schroedinger equation becomes
\begin{align}
i \frac{1}{\beta} \frac{\partial}{\partial \tilde t} \tilde \psi(z,\tilde t)
=&
-\oh
\frac{\partial^2}{ \partial z^2 } \tilde \psi(z,\tilde t)
\nonumber\\
&+
\frac{g}{n} 
z^n \tilde \psi(z,\tilde t)
+
\oh \omega^2
\frac{1}{|\tilde t|^{(n-2) \alpha}}
z^2 \tilde \psi(z,\tilde t)
\nonumber\\
&
+
i
\frac{\alpha}{2 \beta}
\frac{1}{\tilde t}
\left( z \frac{\partial}{\partial z} + \frac{\partial}{\partial z}
z\right) \tilde \psi(z,\tilde t)
,
\end{align}
which is exactly the Schroedinger equation associated
with eq. \eqref{eq:Action_z^n}. 

If $\alpha>0>\beta$ the $z^n$ term is dominating for
$\tilde t\rightarrow \pm \infty$ ($t\rightarrow0^\pm$) as in the
classical motion
then we get a complete set of wave functions as
\begin{equation}
\psi_k(y, t)
\sim_{t\rightarrow 0}
|t|^{ \left| \frac{\alpha}{2 \beta} \right| }
\exp\left(- i E_k \beta
\frac{sgn(t)}{ |t|^{ \left| \frac{1}{\beta} \right| }} 
\right)
\tilde \psi_k(z= |t|^{ \left| \frac{\alpha}{\beta} \right| } y)
,
\label{eq:wave_functions_interacting_model_alpha_positive}
\end{equation}
where $E_k$ it the k-th energy eigenvalue of the effective Hamiltonian
$H_{eff}= \oh p_z^2+ \frac{1}{n} g z^n$
and effective time $t_{eff}=\beta \tilde t$.

The wave functions are normalizable and vanish for $t\rightarrow 0$
allowing for a nice and ``smooth'' crossing of the singularity.
The vanishing of the wave function can be again interpreted as the fact that
the classical particle is spread over all the possible values of $y$.
Since $\left| \frac{\alpha}{2 \beta} \right|= \frac{2 A}{n+2}$
the wave functions vanish (generically) slower than the non interacting
case
and this can be interpreted as the fact that interactions has a better
behavior than the non interacting case.
Better means that classical interacting particle goes to infinity slower
than the free one.

The other case is $\beta>0>\alpha$
as in the classical motion.
In this case 
the $y$ kinetic term  is dominating for
$\tilde t\rightarrow 0^\pm$ ($t\rightarrow0^\pm$).
In fact in this limit the Schroedinger equation is
\begin{align}
i \frac{1}{\beta} \frac{\partial}{\partial \tilde t} \tilde \psi(z,\tilde t)
\sim&
-\oh
\frac{\partial^2}{ \partial z^2 } \tilde \psi(z,\tilde t)
+
i
\frac{\alpha}{2 \beta}
\frac{1}{\tilde t}
\left( z \frac{\partial}{\partial z} + \frac{\partial}{\partial z}
z\right) \tilde \psi(z,\tilde t)
.
\end{align}
Redefining
$\tilde \psi(z,\tilde t) =
e^{i \oh \frac{\alpha}{\beta} \frac{z^2}{\tilde t} }\Psi(z,\tilde t)$
we get
\begin{align}
i \frac{1}{\beta} \frac{\partial}{\partial \tilde t} \Psi(z,\tilde t)
\sim&
-\oh
\frac{\partial^2}{ \partial z^2 } \Psi(z,\tilde t)
-
\oh 
\frac{\alpha (\alpha+1) }{ ( \beta \tilde t )^2}
z^2
\Psi(z,\tilde t)
,
\end{align}
which is the Schroedinger equation derived
from \eqref{eq:Action_z^n_integrated_by_part} and
can be seen as a time dependent harmonic oscillator with
$\Omega_{eff}^2 = - \frac{ \alpha(1+\alpha) }{ t_{eff}^2 }$ (so that
$A_{eff}=-\alpha$ as in the classical case) 
and $t_{eff}=\beta \tilde t$ and therefore it exists as a theory.
In particular we get the leading behavior for the ``ground state''
\begin{align}
\psi(y,t)
=&
| \tilde t|^{-\oh \alpha}
e^{i \oh \frac{\alpha}{\beta} \frac{z^2}{\tilde t} }\Psi(z,\tilde t)
\nonumber\\
&\sim
 |t|^{ - \frac{2\alpha + 1}{\beta} }
 e^{\left[
 \oh i
 \left(
 \frac{2 \alpha + 1}{\beta}
 \frac{sgn(t)}{  |t|^{ \frac{2\alpha + 1}{\beta} } }
 -
 \frac{2\alpha + 1}{ |\beta|^{2\alpha +1}}
 \frac{b_0}{b_1}  |t|^{ - 2 \frac{2\alpha + 1}{\beta} }
 \right)
 y^2
 \right]
 }
 ,
 \label{eq:quantum_wave_a_lt_0}
\end{align}     
where $b_0= c_0 \omega^{A_{eff}}$ and $b_1= c_1 \omega^{1-A_{eff}}$,
i.e. we have reabsorbed the $\omega$ dependence in \eqref{eq:asymp_cX}
into the coefficients which must therefore satisfy an equation
corresponding to \eqref{eq:c0_c1_normalization} without $\omega$.
Finally notice that
$\frac{2\alpha + 1}{\beta} =
-\frac{ 2 n A - (n+2) }{ n+2 }$
so that  when perturbation theory breaks down, i.e.
when $2 A > \frac{n+2}{n-2}$ ($\alpha <-1$)
the wave function vanishes when $t\rightarrow 0$
and the potential unbounded.
Notice that the wave function vanishes when $t\rightarrow 0$
in a wider range of $A$ values, i.e. $2 A > \frac{n+2}{n}$ ($\alpha <-\oh$)
but not all of them implies a
perturbation theory breakdown because the potential is bounded
($-1< \alpha <-\oh$).

\section{Implications for string theory on temporal orbifolds}
\label{sect:string_on_temporal_orbifolds}

All the previous discussion is for the generic Kasner metrics of which
the \BO{} is a peculiar case.
For the \BO{} where $A=\oh$
the QFTs considered do not suffer from any breakdown and
this  is apparently a puzzle because the string on \BO{} has a
divergence.
The solution of this apparent puzzle is that divergences appear in QFT
when higher derivatives interaction terms (induced by massive string
states \cite{Arduino:2020axy})
or non linear sigma model interactions are included.

The reason we did not discuss the quantum mechanical models associated
with these QFTs is that either they suffer from
Ostrogradskii instability or they are not renormalizable.
In any case this is not a limitation
since it is easy seen that we suffer of the same issues as the models discussed.
We have then  a clear explanation of the origin of the divergences in
four point amplitudes .
These divergences are also present in three point amplitudes with
massive states, i.e. in the lowest order of perturbation theory.

This does not mean that gravitational backreaction is not going to
play any role.
In facts in the open string case  when solved the issues at tree level
it may be well reappear to one loop open string
amplitudes.      
This is however not at all obvious since the previous argument on
perturbation theory breakdown applies to closed string as well so the
resolution of the issues at the sphere level with three or four punctures
could suggest the resolution at the annulus level, i.e. the sphere with
two punctures.

Another point worth mentioning is that we have discussed the \BO{}
only and not the \NSO{}.
The reason in this case is technical.
While for the \BO{} and its generalization the Kasner metric we can
reduce the QFT to a quantum mechanical model in the \NSO{} we can only
reduce to a 2d QFT since we need keeping both $x^\pm$.
Nevertheless we expect the same mechanism to be in action for this
case too.

An important point which is worth stressing is that divergences are
present in Lagrangian approach, i.e. in the covariant one where the
time is integrated over but there is no divergence in the light-cone formalism
which is Hamiltonian and where the time is not integrated
\cite{ArduinoFinotelloPesandoLCSFT}.
This is is the same as the previous quantum mechanical models: the
Hamiltonian exists but the perturbation theory does not.
Finally notice that this can be shown explicitly for the \NSO{} which is easily
quantized on the light-cone \cite{ArduinoFinotelloPesandoLCSFT}.
This observation explains also why the matrix model \cite{Craps:2005wd}
is well defined.

Since the problem is essentially Lagrangian this is also an issue for
Witten string field theory and in general for all the covariant formulations.

So we are left with the issue on how treat this divergences.
One possibility is to use the Hamiltonian formalism, for example the
light-cone when available. 
Even if these backgrounds do not possess Poincar\'e symmetry and the
light-cone formalism is well adapted (it is possible to use the
light-cone formalism also in other less obvious
cases \cite{Arduino:2022mir})
one could desire to have a
covariant formulation in this case too then a possible approach is 
\cite{Craps:2013qoa}.
Another possibility is to regularize the theory in some way, for
example non commutativity can do the job \cite{ArduinoPesandoNC}.

Finally let us mention that the way of performing the orbifold
projection in the temporal orbifold cases used in literature
are not on very sound basis since the generators used to write the
orbifold projector are dynamical and they change when interactions are
switched on.
The only clear cut case where this is not the case is the \NSO{} in
light-cone quantization.
If we were to use the proper interacting generators there could also
be some cancellations which could give raise to finite amplitudes.

\section{Conclusions}

First of all let us discuss what the previous computations imply for
QFT and then shortly for string theory since we have discussed string
theory in the previous section.

The first and most important point is that interactions can
drastically change the fate of the fields under a Big Crunch/Big Bang.

Secondly what happens seems to depend on the details of the
interaction, in the models we studied the power of the interaction
$\phi^n$ and the sign of the parameter $\alpha$.

Thirdly the breakdown of the perturbation theory is a breakdown of
Feynman diagram approach, i.e. of the concept of particle.
Obviously this happens because of the spacetime region around the
singularity and excluding this region, i.e. before and after it
the perturbation theory is well defined.
Nevertheless this result rises the question of how to treat
the $S$ matrix in these
backgrounds, in facts the theory exists and  spaces are
asymptotically flat so we could expect to be able to define some kind
of $S$ matrix. Nevertheless it seems that the usual constraints from
unitarity must be revisited since near the singularity the concept of particle
breaks down.

Finally the previous results seem to point to the importance of
minisuperspace  approach and pose the question how to extend it
to string theory.

For the string theory the main result is that, at least, at the tree
level string theory is well for these backgrounds.
Whether divergences from backreaction appear at loop level is by now
unknown also because we have to find a good way of treating the tree level.

\printbibliography[heading=bibintoc]

\appendix
\section{Time dependent harmonic oscillator}
\label{app:time_dep_harm_osc}
We will follows essentially Tseytlin et at \cite{Papadopoulos:2002bg}
which refers to \cite{Lewis:1968tm}
but we will be careful in distinguish between Heisenberg and
Schrodinger representation and this should make things more clear.
For a newer point of view on the problem see also \cite{https://doi.org/10.48550/arxiv.2205.01781}.

As usual we define  operators in Heisenberg picture  as
\begin{equation}
  O_H(t, \tz) = U_S^\dagger(t, \tz) O_S(t) U_S(t, \tz),
\end{equation}
so that we get \HH as ($m>0$)
\begin{equation}
  H_H(t, \tz) =
  \frac{1}{2 m} p_H^2(t, \tz)
  +
  \oh m \Omega^2(t) x_H^2(t, \tz)
,
\end{equation}
where in our case
\begin{equation}
  \Omega^2(t)
  =
  \left( \omega^2 + \frac{A(1-A)}{t^2} \right)
  =
  \left( \omega^2 + \frac{k}{t^2} \right)
.
\end{equation}
We then get the \eom
\begin{align}
  \dot x_H(t, \tz)
  =&
     \frac{1}{m} p_H(t, \tz)
     ,
     \nonumber\\
  \dot p_H(t, \tz)
  =&
  - m\Omega^2(t) x_H(t, \tz)
     ,
\end{align}
with \bcs
\begin{equation}
  x_H(\tz, \tz) = x_S
  ,~~~~
  p_H(\tz, \tz) = p_S
  .
\end{equation}
They imply the second order ODEs
\begin{equation}
  \ddot x_H + \Omega^2 x_H
  =
  \frac{d}{d t}
  \left( \frac{1}{\Omega^2} p_H \right)+ p_H
    =0
    .
\end{equation}

\subsection{Constant Heisenberg creator operator}
We now define the operators $\cA_H(t, \tz)$ using the matrix $\cM(t)$ as
\begin{align}
  \cA_H(t, \tz)
  =&
     \cM(t)
     \cZ_H(t, \tz)
     \nonumber\\
  =
  \left(
  \begin{array}{c}
    A_H(t, \tz) \\ A_H^\dagger(t, \tz) 
  \end{array}
  \right)
  =&
     i
     \left(
     \begin{array}{c c}
       - \dot \cX(t)
       & \frac{1}{m} \cX(t)
       \\
       + \dot \cX^*(t)
       & -\frac{1}{m} \cX^*(t)
     \end{array}
     \right)
         \left(
         \begin{array}{c}
           x_H(t, \tz) \\ p_H(t, \tz) 
         \end{array}
  \right)
  ,  
\end{align}
where $\cX(t)$ is a complex solution\footnote{
As we discuss in appendix \ref{app:complex_classical_solutions} there is a one parameter family of solutions.
}of the classical \eom{} with given
normalization\footnote{    
  Remember that given a second order ODE $\ddot y + a(t) \dot y +
  b(t)=0$
  the Wronskian associated with two solutions $f(t)$ and $g(t)$
  is $W(f,g)= f \dot g - \dot f g$ and it obeys the ODE
  $\dot W + a W=0$ therefore
  $ W= c \exp\left( - \int d t a(t) \right)$ with $c$ a constant.
  In our case $a(t)=0$ and the Wronskian is a constant.
  }
\begin{align}
  \ddot \cX(t) + \Omega^2(t) \cX(t)
  &= 0
    ,
  \nonumber\\
  \cX^* \dot \cX - \cX \dot \cX^*
  &=
    2 i W(\Re \cX, \Im \cX)
    =
    i m
  .
    \label{app:eq:Kr_cX_eqs}
\end{align}
Notice that the previous conditions do not fix completely the
solution. To fix it we need to choose an instantaneous vacuum, see
appendix \ref{app:complex_classical_solutions}.

The previous operators satisfy the relations\footnote
{
  Notice that
  $
  \frac{\partial A_H(t, \tz)}{ \partial t}
  \equiv
  \left(\frac{\partial A(t, \tz)}{ \partial t} \right)_H
  =
  U_S^\dagger \frac{\partial A_S(t, \tz)}{ \partial t} U_S
  $.
  This means that the only reasonable way of computing
  $  \frac{d A_H(t, \tz)}{ d t}$
is to express $A_H$ in terms of operators whose \Scp are time independent.
  }
\begin{align}
  \left[ A_H(t, \tz),  A_H^\dagger(t, \tz) \right]
  &=
    1
    ,
    \nonumber\\
  \frac{d A_H(t, \tz)}{ d t}
  &\equiv
    \left(\frac{\partial A(t, \tz)}{ \partial t} \right)_H
    +
    i \left[ H_H(t, \tz), A_H(t, \tz)\right]
    =
    0
    ,
\end{align}
i.e. the canonical commutation relation and the time independence
relation.

For later use we note that the inverse of $\cM(t)$ is
\begin{equation}
  \cM^{-1}(t)
  =
  \left(
    \begin{array}{c c}
      \frac{1}{m} \cX^*(t) & \frac{1}{m} \cX(t)
      \\
      + \dot \cX^*(t) &  \dot \cX(t)
    \end{array}
  \right)
  .
\end{equation}
   
\subsection{Comparing with the usual harmonic oscillator 1}
The general solution for the $\cX$ equation for the usual harmonic
oscillator is
\begin{equation}
  \cX(t)= \cX_+ e^{i \omega t} + \cX_- e^{-i \omega t}
  ,
\end{equation}
then we can compute the constraint
\begin{align}
  \cX \dot \cX^* - \cX^* \dot \cX
  &= -i m
    \nonumber\\
  &=
    2 i \omega
    \left(
    |\cX_-|^2 - |\cX_+|^2
    \right)
,    
\end{align}
from which we get the solution 
\begin{equation}
  \cX_+ = \sqrt{\frac{m}{2 \omega}} e^{i  \omega (t-\tz)} 
  ,~~~~
  \cX_-=0
  .
\end{equation}
Notice that the constraint fixes $\cX_\pm$ up to a phase that we have
chosen so  that the time invariant Heisenberg operator
\begin{equation}
  A_H(t, \tz)
  =
  \sqrt{\frac{m}{2 \omega}} e^{i  \omega (t-\tz)}
  \left( - i \omega x_H(t,\tz) + \frac{1}{m} p_H(t, \tz) \right)
  ,
\end{equation}
matches the corresponding Schroedinger operator for $t=t_0$. 

\subsection{Hilbert space}
We want to construct the Hilbert space of states to be used in
Heisenberg formalism, i.e. we want states that do no depend on time.

We notice that acting with $U_S$ on the $A_H$ defining  equation we get
\begin{equation}
  \cA_H(t, \tz)
  =
  \cM(t)
  \cZ_H(t, \tz)
  \Longrightarrow
  \cA_S(t)
  =
  \cM(t)
  \cZ_S
  ,
\end{equation}
but because of the \bcs on $\cZ_H$ we can also write
\begin{equation}
    \cA_H(\tz, \tz)
    =
    \cM(\tz)
    \cZ_H(\tz, \tz)
    =
    \cM(\tz)
    \cZ_S
    =\cA_S(\tz)
,
\end{equation}
then because $\cA_H$ is constant we get the basic result
\begin{align}
  \cA_H(t, \tz)
  \equiv&
     U_S^\dagger (t, \tz)  \cA_S(t)   U_S (t, \tz)
          \nonumber\\
  =
     \cA_H(\tz, \tz)
  =&
  \cA_S(\tz)
.
\end{align}

Now we can introduce the ``vacuum'' at time $\tz$ as
\begin{equation}
  A_S(\tz) \ksi 0 \tz
  =0
  ,
  \label{eq:Kr_model_vacuum_t0}
\end{equation}
and build a basis for the Hilbert space which is characterized by time $\tz$ as
\begin{equation}
  \cH_{B \tz}=
  \left\{
    \ksi n \tz = \frac{1}{\sqrt{n!}} A^{\dagger n}_S(\tz) \ksi 0 \tz
  \right\}
  .
\end{equation}

\subsection{Time evolution of basis elements and wave functions 1}
Given any element of the previous basis we can identify it as a
Schroedinger state as
\begin{equation}
  \ksiS n \tz \tz \tz = \ksi n \tz
  ,
\end{equation}
and compute its time evolution as follows.

Let us start with the ``vacuum'', and write
\begin{equation}
  A_S(\tz)   \ksiS 0 \tz \tz \tz
  =
  U_S^\dagger (t, \tz)  A_S(t)  U_S (t, \tz) \ksiS 0 \tz \tz \tz
  =
  U_S^\dagger (t, \tz)  A_S(t)    \ksiS 0 \tz t \tz
  =
  0
  ,
\end{equation}
hence we can determine the time evolution of the $\tz$ vacuum  state
as
\begin{equation}
  A_S(t)    \ksiS 0 \tz t \tz
  =
  0
  ,
\end{equation}
from which follows its wave function up to a time dependent
normalization
\begin{equation}
  \left( - \dot \cX(t) x - \frac{i}{m} \cX(t) \partial_x \right)
  \psi_{0 \{\tz\}}(x, t, \tz)
  =
  0
  \Longrightarrow
  \psi_{0 \{\tz\}}(x, t, \tz)
  =
  \cN(t) e^{i \frac{m}{2} \frac{ \dot \cX(t) }{ \cX(t)} x^2}
  .
\end{equation}
The normalization can be fixed using the Schroedinger equation as
\begin{align}
  i \partial_t   \psi_{0 \{\tz\}}(x, t, \tz)
  =&
     \left[
     i \frac{\dot \cN }{\cN}
     - \frac{m}{2}   \frac{d}{d t}  \left( \frac{\dot \cX}{\cX}      \right)
     \right]
     \psi_{0 \{\tz\}}(x, t, \tz)
     \nonumber\\
  =
  H_S(t) \psi_{0 \{\tz\}}(x, t, \tz)
  =&
     \left\{
     - \frac{1}{2 m}
     \left[ i m \frac{\dot \cX}{\cX}
     -
     m^2 \left( \frac{\dot \cX}{\cX} \right)^2 x^2
     \right]
     +
     \oh m \Omega^2 x^2
     \right\}
     \psi_{0 \{\tz\}}(x, t, \tz)
     ,
\end{align}
and using $\cX$ \eom{} to get
\begin{equation}
  \cN(t)
  =
  \frac{C}{\sqrt{\cX(t)}}
  ,
\end{equation}
with $C$ a constant
which can be fixed requiring the normalization of $     \psi_{0
  \{\tz\}}(x, t, \tz) $ as
\begin{align}
  (     \psi_{0 \{\tz\}}(x, t, \tz),
  \psi_{0 \{\tz\}}(x, t, \tz)
  )
  &=
    \frac{|C|^2}{|\cX|}
    \sqrt{\frac{\pi}{ m \Im\left( \frac{\dot \cX}{\cX}\right) }}
    \nonumber\\
  =&
     |C|^2
     \sqrt{ \frac{2 \pi}{m^2}}
     =
     1
     ,
\end{align}
where we have used $\cX$ normalization and \eom to write
\begin{equation}
  \Im\left( \frac{\dot \cX}{\cX}\right)
  =
  \frac{ \Im\left( {\dot \cX}{\cX^*}\right) } { |\cX|^2}
  =
  \frac{m}{2 |\cX|^2}
  ,
  \label{app:eq:time_dep_harm_osc_normalizability}
\end{equation}
where it is interesting to notice that {\sl the chosen $\cX$ normalization
allows for the convergence of the integral}.
Finally we can write the normalized wave function as
\begin{equation}
  \psi_{0 \{\tz\}}(x, t, \tz)
  =
  \sqrt[4]{\frac{m^2}{2 \pi}}
  \frac{1}{\sqrt{\cX(t)}}
  e^{i \frac{m}{2} \frac{ \dot \cX(t) }{ \cX(t)} x^2}
  .
\label{eq:Kr_psi_n=0}\end{equation}

\subsection{Comparing with the usual harmonic oscillator 2}
Using the results from the previous section and
$\frac{\dot  \cX}{\cX}= i \omega$ we get the harmonic oscillator
ground state wave function
\begin{equation}
  \psi_{0 \{\tz\}}(x, t, \tz)
  =
  \sqrt[4]{\frac{m \omega}{\pi}}
  e^{-i \oh \omega ( t - \tz) }
  e^{- \frac{m}{2} \omega x^2}
  .
\end{equation}

\subsection{Time evolution of basis elements and wave functions 2}
To deal with excited states is better to use a generating function and
therefore we define
\begin{align}
  \kcohsiS z \tz t \tz
  &=
    \sum_{n=0}^\infty \frac{z^n}{\sqrt{n!}} \ksiS n \tz t \tz
    \nonumber\\
  &=
    e^{z A_S^\dagger(t)} \ksiS 0 \tz t \tz
    ,
\end{align}
then we evaluate
\begin{align}
\langle x  \kcohsiS z \tz t \tz
  &=
    \langle x  | U_S(t, \tz) \kcohsi z \tz 
    =    
    \sum_{n=0}^\infty \frac{z^n}{\sqrt{n!}} \psi_{ n\{\tz\}}(x, t, \tz)
    \nonumber\\
  &=
    \langle x  |
    e^{i z \dot \cX^*(t) x_S}
    e^{-i z \frac{1}{m} \cX^*(t) p_S}
    e^{-i \oh z^* \frac{1}{m} \dot \cX^*(t) \cX^*(t)}
    \ksiS 0 \tz t \tz
    \nonumber\\
  =&
     e^{i z \dot \cX^*(t) x
     -i \oh z^* \frac{1}{m} \dot \cX^*(t) \cX^*(t)
     }
    \langle x  |
    e^{- z \frac{1}{m} \cX^*(t) \partial_x}
     \ksiS 0 \tz t \tz
     \nonumber\\
  =&
     \sqrt[4]{\frac{m^2}{2 \pi}}
     \frac{1}{\sqrt{\cX(t)}}
     e^{i \frac{m}{2} \frac{ \dot \cX(t) }{ \cX(t)} x^2
     +
     \frac{m}{\cX(t)} x z
     -
     \oh \frac{\cX^*(t)}{\cX(t)} z^2
     }
     ,
\label{app:eq:generating_function_harm_osc}\end{align}
upon the use of the $\cX$ normalization condition.
It can also be checked that the previous equation satisfy the
Schroedinger equation
\begin{equation}
i \partial_t \langle x  \kcohsiS z \tz t \tz
= \left( - \frac{1}{2 m} \partial^2_x + \oh m \Omega^2(t) x^2\right)
\langle x  \kcohsiS z \tz t \tz
.
\end{equation}

\subsection{Overlaps}
Since we want to check that overlaps are well defined
we need computing
$ \bsi n \tz \ksiS l \tz t \tz$ but it is actually simpler to compute 
\begin{equation}
  \bcohsiS z \tz \tz \tz \kcohsiS w \tz t \tz
  =
  \bcohsi z \tz  U_S(t, \tz) \kcohsi w \tz 
  =
  \sum_{n,l=0}^\infty
  \frac{z^{*n}}{\sqrt{n!}}
  \frac{w^l}{\sqrt{l!}}
  \bsi n \tz \ksiS l \tz t \tz
  ,
\end{equation}
since $\ksi n \tz = \ksiS n \tz \tz \tz$.
Performing the explicit $x $ integral we get
\begin{align}
  \bcohsiS z \tz \tz \tz \kcohsiS w \tz t \tz
  =&
     \sqrt{
     \frac{-i m}{ \dot \cX^*(\tz) \cX(t) - \dot \cX(t) \cX^*(\tz)}
     }
     \nonumber\\
  &
     e^{
     \frac{-i m}{ \dot \cX^*(\tz) \cX(t) - \dot \cX(t) \cX^*(\tz)}
     \left[
     \sqrt{\frac{\cX(t)}{\cX^*(\tz)}} z^*
     +
     \sqrt{\frac{\cX^*(\tz)}{\cX(t)}} w     
     \right]^2
     -
     \oh \frac{\cX(\tz)}{\cX^*(\tz)} z^{*2}
     -
     \oh \frac{\cX^*(t)}{\cX(t)} w^{2}
     }
     .
\end{align}

\subsection{Evolution operator in $x$ space}
{For the same reason as before, i.e. to check the finitness of the
regularized string theory} we need the kernel or the evolution operator
in $x$ space.
We perform the computation using the generating function as follows
\begin{align}
  \langle x_2, t_2 | x_1, t_1 \rangle
  =&
     \langle x_2  | U_S(t_2, t_1) | x_1 \rangle
     =
     \sum_{n=0}^\infty
     \psi_{n\{\tz\}}(x_2, t_2, \tz)
     \psi^*_{n\{\tz\}}(x_1, t_1, \tz)
     \nonumber\\
   &=
     \int \frac{d^2 z}{\pi} e^ {-|z|^2}
     \langle x_2  \kcohsiS z \tz {t_2} \tz
     \bcohsiS z \tz {t_1} \tz    x_1\rangle
     ,
\end{align}
where the $d^2 z$ integral is normalized as
$     \int \frac{d^2 z}{\pi} e^ {-|z|^2}=1 $.

We get
\begin{align}
  \langle x_2, t_2 | x_1, t_1 \rangle
  =&
     \sqrt{
     \frac{-i m^2}{4 \pi \Im\left( \cX(t_2) \cX^*(t_1) \right)}
     }
     \nonumber\\
  &
    e^{
    \frac{i m^2}{4 \Im\left( \cX(t_2) \cX^*(t_1) \right)}
    \left[
    \frac{\cX(t_1)}{\cX(t_2)} x_2^2
    +
    \frac{\cX^*(t_2)}{\cX^*(t_1)} x_1^2
    -
    2 x_1 x_2
    \right]
    +
    i \frac{m}{2}
    \left[
    \frac{\dot \cX(t_2)}{\cX(t_2)} x_2^2
    -
    \frac{\dot \cX^*(t_1)}{\cX^*(t_1)} x_1^2
    \right]
    }
    .
\end{align}

\subsection{Comparing with the usual harmonic oscillator 3}
Using the explicit solution for the harmonic oscillator we get
\begin{equation}
  \Im\left( \cX(t_2) \cX^*(t_1) \right)
  = \frac{m}{2 \omega} \sin \omega (t_2 - t_1)
  ,~~~~
  \frac{\cX(t_1)}{\cX(t_2)}
  = e^{ -i \omega (t_2 - t_1) }
  ,
\end{equation}
then the $x_2^2$ coefficient becomes
\begin{align}
    \frac{i m^2}{4 \Im\left( \cX(t_2) \cX^*(t_1) \right)}
    \frac{\cX(t_1)}{\cX(t_2)}
    +&
    i \frac{m}{2}
    \frac{\dot \cX(t_2)}{\cX(t_2)}
\nonumber\\
&=
    \frac{i m \omega }{ 2 \sin \omega (t_2 - t_1)}
    \left[
      e^{ -i \omega (t_2 - t_1) }
      +
      i \sin \omega (t_2 - t_1)
    \right]
    =
    \frac{ i m \omega}{2}
    \frac{ \cos \omega (t_2 - t_1) }{ \sin \omega (t_2 - t_1)}
    ,
\end{align}
as it should.

\section{Complex classical solution for $L_B$}
\label{app:complex_classical_solutions}
We want to solve the equations \eqref{eq:Kr_cX_eqs}.
One possibility is to use the WKB approach, i.e. the adiabatic vacuum
approach \cite{Fulling:1989nb} and write
\begin{equation}
  \cX(x)= \frac{m}{ 2 W(t) } e^{i \int d t' W(t')}
  ,~~~~
  W^2(t)= \Omega^2(t)+ \delta_1(t) + \frac{\delta_2(t)}{ \Omega^2(t) }
  + O\left( \Omega^{-4} \right)
  ,
\end{equation}
but this approach singles out $\Omega$ as a whole
while for our purposes we are more interested in singling out $\omega$.

\subsection{Perturbative solution for $\cX$ in the small $|\omega t|$ limit}

We want to solve the classical equation with normalization condition
given in \eqref{eq:Kr_cX_eqs} which we repeat here without setting $m=1$
\begin{align}
  \ddot \cX(t) + \Omega^2(t) \cX(t)
  &= 0
    ,
  \nonumber\\
  \cX^* \dot \cX - \cX \dot \cX^*
  &=
    i m
  .
    \label{app:eq:Kr_cX_eqs}
\end{align}
Actually we are interested in the perturbative solution around $t=0$.
This is a second order linear equation and therefore it has two
independent solutions.
For our purpose it is sufficient to consider the following leading
order expansion
\begin{equation}
\cX(t) =
\left\{
\begin{array}{l c}
c_0 (\omega t)^A (1+ O(t^2) ) + c_1 (\omega t)^{1-A} (1+ O(t^2) ) & t>0
\\
\bar c_0 (-\omega t)^A (1+ O(t^2) ) + \bar c_1 (-\omega t)^{1-A} (1+ O(t^2) ) & t<0
\end{array}
\right.
.
\label{app:eq:asymp_cX}
\end{equation}
We allow for different coefficients for $t>0$ and $t<0$ because of the
singularity in the differential equation.
The normalization condition then implies
\begin{equation}
-(2 A - 1) \omega |c_1|^2 \Im\left( \frac{c_0}{c_1} \right)
=
+(2 A - 1) \omega |\bar c_1|^2 \Im\left( \frac{\bar c_0}{\bar c_1} \right)
= - \oh m
.
\label{app:eq:norm_cond}
\end{equation}
The issue to solve is the continuation through the singularity $t=0$.
Since we deal with a classical solution we can expect that it must be
as smooth as possible.
For $A>1$ (for $0<A<1$ both independent solutions vanish for $t=0$ and
therefore we take the solution for $A>1$ as the the solution for this range)
the term $|t|^{1-A}$ is divergent but it is the best we can
do to get a continuous trajectory.
This suggests to set $c_1 = \bar c_1$ and therefore
$c_0 = -\bar c_0$ as consequence of the normalization condition.
Notice that the discontinuity in the coefficient $c_0$ does not make
$\cX$ discontinuous, only $\ddot \cX$ is discontinuous.
We are therefore led to 
\begin{equation}
\cX(t) = c_0 \omega t |\omega t|^{A-1} (1+ O(t^2) ) + c_1 |\omega t|^{1-A} (1+ O(t^2) )
.
\label{app:eq:asymp_cX}
\end{equation}
The general solution of the normalization
condition \eqref{app:eq:norm_cond} reads
\begin{equation}
c_0 = \sqrt{\frac{m}{2 (2 A-1) \omega}} \frac{e^{i\alpha}}{\lambda}
\,e^{i \frac{\pi}{4} }
,~~~~
c_1 = \sqrt{\frac{m}{2 (2 A-1) \omega}} {e^{i\alpha}}{\lambda}
\,e^{-i \frac{\pi}{4} }
, \alpha,\lambda\in\R
,
\end{equation}
where $\alpha$ is a trivial overall phase while $\lambda$ parameterizes
different solutions.
Explicitly we can write the normalized complex classical solution as
\begin{equation}
\cX(t)
=
\sqrt{\frac{m}{2 (2 A-1) \omega}}
e^{i\alpha}
\left(
{\lambda}
e^{-i \frac{\pi}{4} } |t|^{1-A}
+
\frac{e^{+i \frac{\pi}{4} }}{\lambda}  sgn(t)\,|t|^{A}
\right)
(1+O(t^2))
,
\end{equation}
so that
\begin{equation}
\frac{\dot \cX}{\cX}
\sim
\frac{1-A}{t}
+
(2A-1) \frac{c_0}{c_1} \omega |\omega t|^{2(A-1)}
=
\frac{1-A}{t}
+
(2A-1) \frac{1}{\lambda^2} \omega |\omega t|^{2(A-1)}
.
\end{equation}
To understand the role of $\lambda$ we can compute
\begin{equation}
|\psi_0(x,t)|^2
\sim
\frac{1}{|\omega t|^{1-A}}
e^{ - \frac{\omega^{2 A-1}}{\lambda^2} |t|^{2(A-1)} x^2} 
,
\end{equation}
from which we see that $\lambda$ parameterizes the instantaneous
vacuum, in fact for small $\omega t_0$ such that $\Omega(t_0)^2>0$
we can compare with the usual harmonic function probability density
$|\psi_{0 (h.o)}(x,t)|\sim e^{-m \Omega(t_0) x^2}$.

\subsection{Continuation through $t=0$ using a regularized equation}
In the previous section we have given a plausible argument on how to
continue the solution across the $t=0$ singularity based on the
continuity.
We can make this argument more rigorous by looking to the solution with
a regularized $\Omega^2(t)$.
This argument is more rigorous if one is willing to accept that it is
meaningful to regularize $\Omega^2(t)$  as
\begin{equation}
\Omega^2(t) =
\left\{
\begin{array}{l c}
\omega^2+ \frac{k}{t^2} & |t|>\epsilon
\\
\omega^2+ \frac{k}{\epsilon^2} & |t|<\epsilon

\end{array}
\right.
.
\label{app:eq:regularized_Omega}
\end{equation}
We choose $\Omega^2(\epsilon)=\omega^2+ \frac{k}{\epsilon^2}<0$, i.e.
we take  $A>1$ so that
\begin{equation}
|\Omega(\epsilon)|
=
\frac{\sqrt{|k|}}{\epsilon}
-
\oh \frac{\omega^2}{\sqrt{|k|}} \epsilon
+ O(\epsilon^3)
.
\end{equation}  

Obviously we are not adding anything really new to the previous
argument since we are making $\Omega^2(t)$ finite and continuous and
therefore the solution will be finite and continuous across the
singularity and therefore unique.
It is anyhow interesting to see how the discontinuity in the $c_1$
coefficient arises.

The general solution for $|t|<\epsilon$ is
\begin{equation}
\cX(t)
=
c_e \cosh(|\Omega(\epsilon)| t)
+
c_o \sinh(|\Omega(\epsilon)| t)
,
\end{equation}
so that the normalization condition \eqref{app:eq:norm_cond} reads
\begin{equation}
\Im( c_o^* c_e)
= -\oh \frac{m}{|\Omega(\epsilon)|}
,
\end{equation}
whose general solution is
\begin{equation}
c_e = \sqrt{\frac{m}{2 |\Omega(\epsilon)| }} \rho e^{i \beta}
\,e^{-i \frac{\pi}{4}}
,~~~~
c_o = \sqrt{\frac{m}{2 |\Omega(\epsilon)| }}  \frac{e^{i \beta}}{\rho}
\,e^{+i \frac{\pi}{4}}
.
\end{equation}
We can now match the solution at $t=\epsilon$.
Since the solution for $t=\epsilon^+$ diverges as
$\cX(\epsilon^+)\sim \epsilon^{1-A}$
we have either $\rho\rightarrow \infty$ or $\rho\rightarrow 0$.
In the former case we need $\alpha=\beta$ and get
\begin{equation}
\rho
\sim
\frac{ \lambda \sqrt[4]{|k|} \epsilon^{\oh -A}
}{ \cosh \sqrt{|k|}}
,
\end{equation}
and the solution is essentially even since the odd part is suppressed
while in the latter case we need $\alpha=\beta+ \oh \pi$ and get
\begin{equation}
\rho
\sim
\frac{ \lambda \sqrt[4]{|k|} \epsilon^{\oh -A}
}{ \sinh \sqrt{|k|}}
,
\end{equation}
and the solution is essentially odd.

Letting $A\rightarrow 1^+$, i.e. $|k|\rightarrow 0$ such that
$\frac{|k|}{\epsilon^2}$ is kept constant and bigger than $\omega^2$
we get the usual harmonic oscillator with .....
Then only in the even case $\rho$ has a finite limit while in the odd
case
$\rho\sim \epsilon$.

\section{WKB analysis of $L_B$}
\label{app:WKB_analysis_time_dep_ham_osc}
but we will use the WKB expansion and write
$\psi(x,t) = e^{i S(x,t) }$ so that we want to solve the equation
\begin{equation}
   \partial_t S(x,t)
  +\oh (\partial_x S(x,t))^2
  + \oh \left( \omega^2 + \frac{k}{t^2} \right)
  - i \oh \partial_x^2 S(x,t)
  = 0
  .
\end{equation}
In the limit $t\rightarrow 0$ we can try to write
\begin{equation}
  S(x,t)
  =
  \theta(t) t^{a_{(+)}} s_{(+)0}(x) (1 + o(1))
  +
  \theta(-t) (-t)^{a_{(-)}} s_{(-)0}(x) (1 + o(1))
\end{equation}
and fix $a_{(\pm)}$.
Notice that we allow for a discontinuity in $S$ at $t=0$ since
$p_x \psi(x,t) = \partial_x S \psi(x,t)$ and the momentum can be
discontinuous due to the infinite force.

At the leading order in $t$ we get 
\begin{equation}
  a \frac{|t|^a}{t} s_0(x)
  +\oh |t|^{2 a} (s'_0(x))^2
  +\oh \frac{k}{t^2} x^2
  -i \oh |t|^a s''_0(x)
  \sim0
  ,
\end{equation}
where we have dropped the subscript $(\pm)$ since the equation is the
same for both cases.
The unique solution is $a=-1$.
Then we are left with
\begin{equation}
  (s'_0(x))^2  - 2 s_0(x) + k =0 
  .
\end{equation}
This is a special case of Chrystal's equation\footnote{
  Chrystal's equation reads
  \begin{equation*}
    \dot y ^2 + A t\dot y+ B y + C x^2=0
    .
  \end{equation*}
  The general solution is
  \begin{equation*}
    x  \frac{(z-a)^{a/(a-b)}}{(z-b)^{b/(a-b)}} = k,
    ~~~~
    4 B y=(A^2-4C-z^2) x^2
    ,
  \end{equation*}
  and $a,\ b = \pm  \left[ B +\sqrt{(2A+B)^2 - 16C} \right]/2$.
}.
The most singular and easiest solution is
\begin{equation}
  S_0(x) = \oh \alpha x^2
  ,~~~~
  \alpha^2-\alpha+k=0 \Rightarrow \alpha\in\{A,1-A\}
  .
\end{equation}
Since there are two solutions for $\alpha$ it is still possible that
$s_{(+)0}(x)$ differs from $s_{(-)0}(x)$ but it turns out that they
are the same since in order to avoid
singularities at $x=0$ for $t\ne0$ in $S$ we need choosing the lowest
$\alpha$ solution.

So one could think of setting up an expansion like
$S= \frac{1}{t} s_0(x) + s_1(x) + t s_2(x) +O(t^2)$.
This is possible but does not give the right answer.
The equation is non linear and therefore we cannot add solutions 
hence we must check whether there exist subdominant expansions.
Let us therefore write
\begin{align}
  S(x,t)
  =&
  \frac{1}{t} s_{0}(x) (1 + O(t))
  +  t^{\alpha b} \bs{1}_{0}(x) (1 + O(t))
  ,
\end{align}
and try to fix $b$ such that $\Re(\alpha b)>-1$ and the added term is
actually subdominant.
The equation for $\bs{1}_{0}(x)$ turns out to be
\begin{equation}
  2 s_0' \bs{1}_0'+ 2 \alpha b \bs{1}_0=0
  ,
\end{equation}
which has solution
\begin{equation}
  \bs{1}_0(x) = \bc{1}_0 |x|^{-b}
  ,
\end{equation}
since we do not want singularities in $x$ we need
$\Re b <0$.
So it seems that any $b$ which satisfies the previous constraint
can do but again the request of singularities
in $x$ in higher order terms forces $b=-1$.
It follows therefore that $\alpha<0$ since $\alpha$ is real.

Finally we can set up the perturbative expansion as
\begin{align}
  S(x,t)
  =&
  \frac{1}{t} s_{0}(x) + s_1(x) + t s_2(x) + O(t^2)
  \nonumber\\
  &+ log(|t|) \hat s_1
  \nonumber\\
  &+
  |t|^{\alpha b} \bs{1}_{ 0}(x)
  +  |t|^{\alpha b} t \bs{1}_{(+) 1}(x)
  +  O(t^{\alpha b+2}) 
  \nonumber\\
  &+  |t|^{2 \alpha b} \bs{2}_{0}(x)
  +  |t|^{2 \alpha b} t \bs{2}_{1}(x)
  +  O(t^{2\alpha b+2})
  \nonumber\\
  &+ \dots
  ,
\end{align}
where we added a further logarithmic contribution with {\sl constant}
coefficient $\hat s_1 = (\oh i +\delta ) \alpha$ which is necessary
for the absence of singularities in $x=0$ from $s_1$ and added  double
infinite series with power $t^{n b}$ since as soon as we add $t^b$
we get a term with power $t^{2 b}$ from $(\partial_x S)^2$.
In the case of non integer power we need paying attention to the
definitions of $\bs{n}_m$ in order to get equations which do not depend
on the sign of $t$ therefore we write $|t|^{n b} t^m$.
Finally notice that we need not only $\Re(b)>-1$ but $\Re(b)\ge 0$ so
that all added terms are subdominant.

We now get the equations
\begin{align}
|t|^{\alpha b}/t:&&
2 s_0' \bs{1}_0' + 2 \alpha b \bs{1}_0  =0
\nonumber\\
|t|^{2 \alpha b}/t:&&
2 s_0' \bs{2}_0' + 4 \alpha b \bs{2}_0  =0
\nonumber\\
  t^{-1}:&&
  2 s_0' s_1'-i s_0'' +2 \hat s_1 = 0
\nonumber\\
|t|^{\alpha b}:&&
2 s_0' \bs{1}_1' + 2(\alpha b+1) \bs{1}_2 -i \bs{1}_0'' + 2 s_1' \bs{1}_0' +
= 0
\nonumber\\
|t|^{2 \alpha b}:&&
2 s_0' \bs{2}_1' + 2(2 \alpha b+1) \bs{2}_1
-i \bs{2}_0'' +2s_1' \bs{2}_0' + (\bs{1}_0')^2
=0
\nonumber\\
t^{0}:&&
2 s_0' s_2' + 2 s_2 -i s_1'' + (s_1')^2 + \omega^2 x^2 =0
.
\end{align}
The solution for $\bs{1}_0$ and $\bs{2}_0$ read
\begin{equation}
    \bs{1}_0(x) = \alpha \bc{1}_0 |x|^{-b}
    ,~~~~
    \bs{2}_0(x) = \alpha \bc{2}_0 |x|^{-2 b}
,
\end{equation}
from which one can easily guess the solution for all $\bs{n}_0$.
The solution for $s_1$, $\bs{1}_1$ and $\bs{2}_1$ read
\begin{align}
  s_1(x) =& \alpha c_1 - \delta log |x|,
  \nonumber\\
  \bs{1}_1(x) =&
  \frac{\alpha b \bc{1}_0 ( 2 \delta -i b - i ) }
       {2 (2 \alpha -1)}
       |x| ^{-b -2  }
       +
       \alpha \bc{1}_1 |x| ^{-b - \frac{1}{\alpha} }
  \nonumber\\
  \bs{2}_1(x) =&
  \frac{\alpha b [ \bc{2}_0 ( 4 \delta - 4 i b - 2 i ) + \alpha b \bc{1}_0^2]}
       {2 (2 \alpha -1)}
       |x| ^{-2 b -2  }
       +
       \alpha \bc{2}_1 |x| ^{- 2 b - \frac{1}{\alpha} }
       .
\end{align}
Finally we get also
\begin{align}
  s_2(x) =&
  - \frac{\omega^2}{2(2\alpha+1)} x^2
  + \frac{\delta (\delta-i)}{2(2\alpha+1)} \frac{1}{x^2}
  - \alpha c_2 |x|^{-\frac{1}{\alpha}}
  .
\end{align}
Assembling all pieces in order to discuss the constraints on the
constants we get
\begin{align}
  S(x,t)
  =&
  \frac{1}{t} \left[ \oh \alpha x^2 \right]
  + \log |t| \left[ \alpha ( \frac{i}{2} + \delta) \right]
  + \left[ - \delta \log|x| \right]
  \nonumber\\
  &
  +
  t
  \left[
  - \frac{\omega^2}{2(2\alpha+1)} x^2
  + \frac{\delta (\delta-i)}{2(2\alpha+1)} \frac{1}{x^2}
  - \alpha c_2 |x|^{-\frac{1}{\alpha}}
  \right]
  \nonumber\\
  &
  + |t| ^{\alpha b}
  \left[
    \alpha \bc{1}_0 |x|^{-b}
    \right]
  + |t| ^{\alpha b} t
  \left[
      \frac{\alpha b \bc{1}_0 ( 2 \delta -i b - i ) }
       {2 (2 \alpha -1)}
       |x| ^{-b -2  }
       +
       \alpha \bc{1}_1 |x| ^{-b - \frac{1}{\alpha} }
    \right]
  \nonumber\\
  &
  + |t| ^{2 \alpha b}
  \left[
    \alpha \bc{2}_0 |x|^{-2 b}
    \right]
  +
  |t| ^{2 \alpha b} t
  \left[
  \frac{\alpha b [ \bc{2}_0 ( 4 \delta - 4 i b - 2 i ) + \alpha b \bc{1}_0^2]}
       {2 (2 \alpha -1)}
       |x| ^{-2 b -2  }
       +
       \alpha \bc{2}_1 |x| ^{- 2 b - \frac{1}{\alpha} }
    \right]
\end{align}
The absence of singularities in $x=0$ in the $O(t)$ term
implies $\delta=0$ or $\delta=i$.

Finally when $A>1$ ad $\alpha=1-A<0$ we get
\begin{equation}
  \psi(x,t)
  =
  N |t|^{\oh (A-1)}
  e^{ i \left[
      - \frac{A-1}{4} \frac{1}{t} x^2
      +t \frac{\omega^2}{2(A-1)} x^2
      \right] }
  e^{ - i (A-1) \bc{1}_0 |t|^{A-1} x
    +\oh \bc{1}_0^2 |t|^{2 (A-1)} x^2
    }
  .
\end{equation}
As long as we take $\Re( \bc{1}_0^2 )<0$ this expression is consistent
since the normalization $N$ is a constant and independent on $x$ and
$t$ as it follows from
\begin{align}
  \int_{-\infty}^\infty d x |\psi(x,t)|^2
  =&
  \int_{-\infty}^\infty d x 
  |N|^2 |t|^{ (A-1)}
  e^{
    2 \Re( \bc{1}_0^2 ) |t|^{2 (A-1)} x^2
    -
    2 (A-1) \Im( \bc{1}_0 ) |t|^{(A-1)} x
    }
  \nonumber\\
  &=
  |N|^2
  \sqrt{\frac{ 1 }{ -2 \Re( \bc{1}_0^2 )} }
  e^{ (1-A)^2  \frac{ (\Im \bc{1}_0 )^2 }{ -2 \Re( \bc{1}_0^2 ) } }
    .
\end{align}
The physical meaning  of the vanishing of the wave function for $t=0$ is
that the particle is diffused uniformly on the entire real axis $x$.

\end{document}